\documentstyle[emulateapj]{article}

\slugcomment{Submitted to the Ap. J. Supp. Series 21 Dec 1998, Accepted 4 Apr 1999}

\lefthead{Hora, Latter, \& Deutsch}
\righthead{Near-IR PN Spectral Survey}

\def\h2{H$_2$}
\def\s1{$v = 1 \to 0$ S(1)}
\def\Lya{Ly$\alpha$~}

\def\about{$\approx$}

\def\Pab{Pa$\beta$}
\def\Brg{Br$\gamma$}

\def\bd30{BD+30$^{\circ}$3639}
\def\irc10420{IRC+10$^{\circ}$420}
\def\etal{et al.\ }
\def\eg{e.g.,\ }

\def\cm3{cm$^{-3}$}
\def\kcm3{{\rm cm}^3~{\rm s}^{-1}}
\def\fcm3{{\rm cm}^{-3}}

\begin{document}

\title{Investigating the Near-Infrared Properties of Planetary 
Nebulae \\II.\ Medium Resolution Spectra}
\author{Joseph L. Hora}
\affil{Harvard-Smithsonian Center for Astrophysics, 60 Garden Street  MS/65, 
Cambridge, MA 02138-1516; jhora@cfa.harvard.edu}
\author{William B. Latter}
\affil{SIRTF Science Center/Infrared Processing and Analysis Center,
California Institute of Technology, MS 310-6, Pasadena, CA  91125; 
latter@ipac.caltech.edu}
\author{Lynne K. Deutsch}
\affil{Astronomy Department, Boston University, 725 Commonwealth Avenue, Boston, MA 02215; deutschl@bu.edu}

\begin{abstract}

We present medium-resolution (R $\sim$ 700) near-infrared ($\lambda =
1 - 2.5$ \micron) spectra of a sample of planetary nebulae (PNe).  A
narrow slit was used which sampled discrete locations within the
nebulae; observations were obtained at one or more positions in the 41
objects included in the survey.  The PN spectra fall into one of four
general categories: \ion{H}{1} emission line-dominated PNe, \ion{H}{1}
and \h2\ emission line PNe, \h2\ emission line-dominated PNe, and
continuum-dominated PNe.  These categories correlate with
morphological type, with the elliptical PNe falling into the first
group, and the bipolar PNe primarily in the \h2\ and continuum
emission groups.  The categories also correlate with C/O ratio, with
the O-rich objects generally falling into the first group and the
C-rich objects in the other groups.  Other spectral features were
observed in all categories, such as continuum emission from the
central star, C$_2$, CN, and CO emission, and warm dust continuum
emission towards the long wavelength end of the spectra.

Molecular hydrogen was detected for the first time in four PNe.  An
excitation analysis was performed using the \h2\ line ratios for all
of the PN spectra in the survey where a sufficient number of lines
were observed.  From the near-infrared spectrum, we determined an
ortho-to-para ratio, the rotational and vibrational excitation
temperatures, and the dominant excitation mechanism of the \h2\ for
many objects surveyed.  One unexpected result from this analysis is
that the \h2\ is excited by absorption of ultraviolet photons in most
of the PNe surveyed, although for several PNe in our survey
collisional excitation in moderate velocity shocks plays an
important role.  The correlation between bipolar morphology and \h2\
emission has been strengthened with the new detections of \h2\ in this
survey. We discuss the role of winds and photons to the excitation of
\h2\ in PNe, and consider some implications to the utility of \h2\ as
a nebular diagnostic and to our understanding of PNe structure and
evolution.

\end{abstract}


\section{Introduction}

This is the second of two papers describing the results of surveys
examining the properties of planetary nebulae (PNe) as observed in the
near-infrared ($\lambda = 1 - 2.5$ \micron).  The first paper (Latter
et al.\ 1995; hereafter Paper I) presented an infrared imaging survey;
here we present the results of a near-infrared spectral survey.

There are several reasons why knowledge of the near-infrared (near-IR)
characteristics of PNe is important, as described in Paper I.  In
order to interpret the imaging results and to learn more about the
physical conditions in the nebulae, the spectra of these objects must
be examined to understand the processes responsible for the
emission. There are many emission lines present in the 1 -- 2.5
\micron\ spectral region, most notably those due to recombination
lines of \ion{H}{1}, and lines of vibrationally excited \h2.  Also
present are atomic lines of \ion{He}{1} and [\ion{Fe}{2}], and
emission from other molecular species such as CO and C$_2$.  These
lines can act as diagnostic tools to probe the physical conditions
inside the nebula, sampling different regions and ranges of
temperature, density, and excitation than is seen by observing the
optical line emission.  Some PNe also exhibit in the near-IR strong
continuum emission from hot dust.  This emission becomes significant
longward of $\lambda = 2$ \micron\ in many of the PNe, requiring
near-IR spectroscopy to detect it and to differentiate between line
and continuum emission sources.  Finally, the lower optical depth of
the PNe in the IR as compared to optical wavelengths allows us to
potentially see into regions of the nebula that are obscured by dust.

There have been several previous surveys that have explored the
properties of PNe in the infrared.  Early spectroscopic and
photometric surveys (e.g., Gillett, Merrill, \& Stein 1972; Cohen \&
Barlow 1974) determined that there was an excess of IR emission over
what was expected from reflected continuum emission from the central
star.  Other photometric surveys in the following years (Whitelock
1985; Persi et al.\ 1987) determined the primary sources of IR
emission to be stellar continuum, thermal dust emission, and thermal
\& line emission from the nebula itself.  The near-IR color
characteristics of most PNe are unique and can be used to identify new
PNe and post-AGB objects (Garcia-Lario et al.\ 1990).

Recently there have been more detailed spectral observations of PNe in
the near-IR.  Hrivnak, Kwok, \& Geballe (1994) surveyed a set of
proto-PNe in the H and K bands.  In these objects the \ion{H}{1}
Brackett lines were observed in absorption, and most objects had CO
absorption or emission, indicating recent mass loss events.  Rudy et
al.\ (1992, and references therein) and Kelly \& Latter (1995) have
surveyed several PNe and proto-PNe in the $\lambda = 0.5 - 1.3$
\micron\ range.  Dinerstein \& Crawford (1998) have completed a survey
of a set of PNe in the K-band, focusing on excitation of molecular
hydrogen.

There are several unique aspects of the survey results presented here
that were made possible by the KSPEC spectrograph (see the instrument
description below). First, because of KSPEC's high sensitivity and
simultaneous sampling of the full near-infrared spectral range, we
were able to obtain data on a comparatively large number of objects
(41) in a short period of time.  Using the relatively narrow and short
slit of the spectrograph, we sampled different regions of the PNe to
examine the emission throughout the nebula.  In most of the other
surveys described above, larger beams were used that included much or
all of the object. Another aspect of the data presented here is
because of the cross-dispersed design of the instrument, the entire
$\lambda = 1.1 - 2.5$ \micron\ range is obtained at once, eliminating
the possibility of telescope pointing errors or other fluctuations
affecting the relative line strengths in the spectra.  Finally, the
slit-viewing detector allowed precise positioning and guiding, so the
region of the PN being observed was well known for each spectrum.  The
PNe observed in this survey were chosen to overlap with the near-IR
imaging survey (Paper I), along with several other optically-bright
PNe and unresolved objects that were not included in the imaging
survey.

\section{Observations and Data Reduction}

The observations were performed on several runs during the period 1992
October through 1994 September at the University of Hawaii 2.2m telescope
on Mauna Kea, using the near-IR KSPEC spectrograph (Hodapp et al.\ 1994). 
KSPEC is a $\lambda = 1 - 2.5$ \micron\ cross-dispersed spectrograph that
has a separate slit-viewing IR array for acquiring the source and guiding. 
The full spectral range is obtained in a single exposure, resulting in
accurate relative line measurements and highly efficient data acquisition.
The diffraction orders are well-matched to the atmospheric transmission
windows, with the K band in 3rd order (1.9 -- 2.5 \micron), H in 4th order
(1.45 -- 1.8 \micron) and J in 5th order (1.15 -- 1.32 \micron), with a
resolution R $\sim$ 700.  The
1$\times$6 arcsecond slit provides a small aperture that was used to
sample different spatial regions of the nebula to search for spectral
variations. 

Table 1 lists the nebulae observed and details of the observations.
Integration times for each exposure ranged from a few seconds for the
extremely bright sources to 5 minutes for faint sources.  The off-axis
guider was used to keep a consistent on-source slit position. Multiple
``Fowler'' sampling was used to reduce the read noise. The number of
samples for a particular integration ranged from 4 to 16, with more
samples used for the longer integration times.  The spectra were reduced
using IRAF; the extraction and processing of the spectral data were done
using the functions in the noao.twodspec and noao.onedspec packages.
Alternating source and sky integrations of the same length were taken and
differenced to remove sky and telescope background flux.  Dome flats were
used to correct for pixel-to-pixel gain variations in the array.  Stars of
known spectral type (either G0 or A0) were observed at the same airmass as
the nebulae immediately before and after the PNe observations and were
used for correction of the instrumental response and sky transmission. 
Individual lines were removed from the stellar spectra, and then
normalized using a blackbody function of T = 5920 K for the G0 stars and T
= 10800 K for the A0 standards.  The spectra were wavelength-calibrated
using observations of an Argon reference lamp.  The wavelength values used
for lines greater than 1.1 \micron\ are from Rao et al.\ (1966); for lines
less than 1.1 \micron, the wavelengths were taken from Wiese, Smith, \&
Miles (1969) and corrected to vacuum wavelengths.  The average 1$\sigma$
uncertainty in the measured wavelengths of the lines is about 5 \AA. 

Infrared photometric standard stars were observed in the same way as the
PNe and used to flux calibrate the spectra.  Absolute calibration is
difficult with these spectra because not all of the flux from the star
enters the narrow slit during a single integration, and the amount differs
for each integration depending on how well the star is centered on the
slit (the seeing at 2 \micron\ during typical observations was 0\farcs5 --
1\farcs0).  The amount of light lost was estimated by the following
method: the full width at half maximum (FWHM) brightness of the standard
star was measured along the spatial direction of the slit, in the spectrum
with maximum flux for that star.  It was then assumed that the point
spread function (PSF) is well represented by a two-dimensional Gaussian
distribution with the measured FWHM, and the amount of flux falling
outside of the slit was then calculated.  This was typically 20 -- 30\% of
the total light for a single integration.  The calibration for each star
was corrected by this factor, along with corrections for airmass. 
Comparing results from different standard stars taken throughout the night
indicated that this method is accurate to approximately 20\%.  For the
observations of the PNe, no correction was applied for the slit width or
length. The
length along the slit of the extracted regions for the three bands were
2\farcs0 (J), 3\farcs5 (H), and 4\farcs0 (K).  

\section{Results }

The spectra are presented in Figures 1 -- 32.  One to three PNe spectra
are plotted in each figure.  
Tables 2 through 7 list the line identifications and extracted fluxes with
uncertainties for the spectra shown in the Figures. 
The PN in this section are listed in Tables 2, 3, and 4.
There are a number of features not identified (indicated by question
marks in the tables). These features tend to appear above the
$\sim3\sigma$ level and must be considered real, though confirming spectra
would be valuable. Our search for possible identifications for these lines
has been careful, but perhaps not exhaustive.  In addition to the
relatively low S/N of these lines, the moderate wavelength resolution is
not sufficient to differentiate between the several possible
identifications for each line. 

The PNe spectra are separated into four groups that share common
characteristics.  These are \ion{H}{1} recombination line-dominated,
\ion{H}{1} recombination line + \h2\ emission, \h2\ -- dominated, and
continuum-dominated.  A fifth group of objects is included that contains
two objects that were at one time classified as PNe, but are now generally
regarded as being \ion{H}{2} regions (M 1--78 and K 4--45; Acker et al.\
1992). We do not discuss these objects further, but include them for
comparison. Within each group, the NGC objects are listed first, followed
by the remaining PNe in alphanumeric order. The morphological
classifications are given according to Balick (1987), unless otherwise
noted. 

\subsection{\ion{H}{1} - line dominated}

The line emission in these PNe is dominated by lines of \ion{H}{1} and
\ion{He}{1}.  In the J-band, the Paschen $\beta$ (\Pab) line is the most
intense, with contributions from lines of \ion{He}{1}, [\ion{Fe}{2}], and
\ion{O}{1}.  In the H band, the Brackett series of \ion{H}{1} dominate,
with \ion{He}{1} emission at 1.7002 \micron\ and [\ion{Fe}{2}] emission at 1.6440
\micron\  present in some PNe.  In the K band, the
brightest line is usually Brackett $\gamma$ (\Brg), with strong lines of
\ion{He}{1} at 2.058 and 2.112 \micron. When the \ion{H}{1} lines are
strong enough, one begins to see the Pfund series lines starting near 2.35
\micron\ where they are just beginning to be separated at this resolution. 
There are also two unidentified lines at 2.199 and 2.287
\micron\ (Geballe et al.\ 1991) that appear in several PNe in this
category.    A few of the spectra
shown here have contributions from central star continuum flux that is
larger towards shorter wavelengths, or warm dust continuum which is 
stronger at longer wavelengths. 

\subsubsection{NGC 1535}

NGC 1535 is classified as early round, and its near-IR spectrum is
dominated by emission lines of \ion{H}{1}.  It has a bright ionized
shell of emission which is surrounded by a fainter halo (e.g., see
Schwarz, Corradi, \& Melnick 1992).  Near-IR images were presented in
Paper I.  This PN has previously been observed to have \h2\ lines in
absorption in the far-UV (Bowers et al.\ 1995).  Recent observations
by Luhman et al.\ (1997) failed to detect \h2\ in emission in the \s1
line.  They attributed the earlier detection of absorption at shorter
wavelengths to the interstellar medium, or a region in the PN itself of
a much smaller size than the ionized zone.

The spectrum of NGC 1535 shown in Figure 1 was obtained at a position
centered on the brightest part of the ring directly west (W) of the
central star.  We also fail to detect the \h2\ emission in the \s1
line, at a 1 $\sigma$ level of about $5\times10^{-17}$ ergs cm$^{-2}$
s$^{-1}$ \AA$^{-1}$.  There is some indication of emission from \h2\ in the $v
= 1\to 0$ Q(1) and $v=1\to 0$ Q(3) lines at the long wavelength end of
the spectrum.  However, the spectrum is noisier in this region and 
there is confusion with the Pfund-series \ion{H}{1} lines, therefore the \h2\
lines are not detected above the 3 $\sigma$ level.

\subsubsection{NGC 2022}

NGC 2022 is an early elliptical PN, but is morphologically very
similar to NGC 1535 in optical images (e.g., Schwarz et al.\
1992). The main difference between the two is a different relative
outer halo size as compared to the inner ring (the halo is relatively
smaller in NGC 2022).  Zhang \& Kwok (1998) also find similar
parameters with their morphological fits of these two PNe. Near-IR
images of this PN were presented in Paper I.

NGC 2022 is also spectrally similar to NGC 1535, as seen in Figure
1. The spectrum of NGC 2022 was taken centered on the ring directly
east (E) of the central star.  The dominant emission lines are those
of atomic hydrogen.

\subsubsection{NGC 2392}

The PN NGC 2392 (the ``Eskimo nebula'') is another double-shell nebula;
however, this PN has a significant amount of structure in the inner ring
and outer shell.  Spectrophotometric (Barker 1991) and kinematic studies
(Reay, Atherton, \& Taylor 1983; O'Dell, Weiner, \& Chu 1990) that have
been carried out with optical imaging and spectroscopy have revealed the
abundances and ionization states and velocities of the various
components.  Near-IR images of this PN were presented in Paper I. 

The spectrum of NGC 2392 in Figure 1 was taken centered on the brightest
part of the ring directly E of the central star.  This third early-type
round PN differs from the other two in Figure 1 primarily from the bright
\ion{He}{1} line at 2.058 \micron, and the [\ion{Fe}{2}] lines in the J
and H bands of the spectrum. 

\subsubsection{NGC 3242}

NGC 3242 is an early elliptical with several interesting morphological
features.  In addition to the bright elliptical ring, there are several
filaments and knots of emission in the central region, and two ansae that
are placed roughly along the major axis of the elliptical emission.  Also,
there is a larger faint halo that envelopes the inner structure. 

Spectra acquired at three different positions on the nebula are shown in
Figure 2, on the SE knot (NGC 3242SE), on the E section of the bright ring
(NGC 3242E), and on the SW halo (NGC 3242H).  The spectra are similar in
all locations; the bright \ion{H}{1} lines are present in all positions,
along with stellar continuum at the shorter wavelengths.  One difference
is that the \ion{He}{2} line at 2.189 \micron\ are much
brighter in the E ring than in the SE knot or halo position. 

\subsubsection{NGC 6210}

NGC 6210 is a fairly compact PN with a core -- halo morphology similar to
other ellipticals.  Phillips \& Cuesta (1996) performed a visible
wavelength spectroscopic study that revealed a complex velocity structure
which suggests multiple shells and possibly ``jets'' at various position
angles.  The near-IR image presented in Paper I does not reveal much of
the structure.  This PN is one of several in which Geballe et al.\ (1991)
detected unidentified emission at 2.286 \micron\ (but not at 2.199 \micron). 

Three spectra are presented for this PN, and shown in Figure 3. They were
acquired with the slit centered on the core (Core), 1\arcsec\ east (E1),
and 3\arcsec\ east (E3).  The core position shows a contribution from
stellar continuum in the $\lambda = 1 - 1.8$ \micron\ region that
decreases successively in the E1 and E3 positions.  The unidentified feature at
2.287 \micron\ is detected at the E1 position, but not the 2.199 line,
similar to the findings of Geballe et al.\ (1991). 

\subsubsection{NGC 6543}

NGC 6543 (the ``Cat's Eye'') has a complex morphology, with a high degree
of symmetry. Recent HST imaging (Harrington \& Borkowski 1994) has shown
more clearly the structure of rings, shock fronts, jets, and fast,
low-ionization emission-line regions (FLIERS) present in this PN. 

The spectrum shown in Figure 4 was taken at the position S of the central
star and slightly E, where the two emission arcs cross and create a local
emission maximum (see Paper I, Figure 6a).  The spectra is similar to the other PN in
the \ion{H}{1}/\ion{He}{1} - dominated class, and also show the
unidentified lines at 2.199 and 2.286 \micron. 

\subsubsection{NGC 6572}

This young PN has a bipolar morphology in the near-IR, with its major axis
in the N-S direction, and a bright ring structure closer to the central
star (Paper I).  The near-IR spectrum from 0.77 to 1.33 \micron\ was
measured by Rudy et al.\ (1991), and the UV and optical spectrum was
recently obtained by Hyung, Aller, \& Feibelman (1994), who found evidence
of variability. 

Figure 5 shows two slit positions on the PN, one on the nebula center,
and one on the brightest location in the E lobe of the PN.  Both the
core and E lobe spectra show a strong contribution from stellar
continuum flux.  There is strong \ion{H}{1} and \ion{He}{1} line
emission, and relatively strong unidentified line emission at 2.199 and
2.286 \micron.

\subsubsection{NGC 6790}

This PN has been shown by radio continuum observations to be an
elliptically-shaped nebula with a diameter of roughly an arcsecond
(Aaquist \& Kwok 1990).  Aller, Hyung, \& Feibelman (1996) obtained UV and
optical spectra and suggest that NGC 6790 is a relatively young object,
slightly more evolved than Hb 12.  Kelly \& Latter (1995) obtained a 0.9
-- 1.3 \micron\ spectrum and find similarities to Hb 12 and AFGL 618. 

The spectrum of NGC 6790 in Figure 4 shows a stellar contribution because
the slit was centered on the object, and includes the star as well as the
nebular emission that is typical of this class of PN. 

\subsubsection{NGC 6803}

This compact elliptical PN has a uniformly bright disk with no
apparent structure in the optical, and is surrounded by a fainter halo
about twice the size of the bright shell (Schwarz et al.\ 1992). In the
KSPEC imaging channel, however, the PN was seen to be double-lobed,
with the lobes on the minor axis of the bright elliptical
region. Spectra were obtained at two positions, one centered on the E
lobe, and the other position 5 arcseconds NW in the halo region.  The
lobe spectrum shows bright \ion{H}{1} and \ion{He}{1} lines, with a
stellar continuum contribution out to about 1.5 \micron. There is also
unidentified line emission at 2.286 \micron.  The halo spectrum is
quite different, with weak \Pab\ and continuum emission, and is
probably reflected star and nebular emission.

\subsubsection{NGC 6826}

This elliptical PN is morphologically similar to NGC 3242, with a bright
elliptical inner ionized ring, ansae along the major axis, and a fainter
halo that envelopes the system (Balick 1987). The I-band image in Paper I
shows evidence for a shell between the inner bright ring and the outer
halo, and at the same radial distance as the ansae. 

The spectra are shown in Figure 7.  The bright core is dominated by
stellar continuum emission.  The other positions show strong 
\ion{H}{1} line emission.  The spectrum labeled SW Lobe was taken on the inner
bright ring directly SW of the central star.  The SW Halo position was
taken in the halo midway between the bright ring and the outer edge of the
PN.  The lobe and halo emission is similar, except for relatively brighter
lines of \ion{He}{1} at 1.7002 \micron\ in the lobe.  There is also some
continuum emission at the short wavelength end of the nebular positions,
which is probably scattered light from the central star. 

\subsubsection{NGC 7009}

NGC 7009 (the ``Saturn'' nebula) is an elliptical with an interesting
twisted symmetry in its shell and in the various filaments and knots of
emission.  Balick et al.\ (1998) recently published HST images that show
the ``microstructures'' in this PN.  The images show that the inner knots
are actually groups of FLIERs, and jets in [\ion{N}{2}] are seen that
terminate at the tips of the nebula. 

Two positions were sampled, one in the halo region on the W edge of the
PN, and one in the N part of the nebula.  Both positions show a
stellar continuum emission contribution from the central star. The
halo emission is similar to the spectrum taken on the north edge of the
PN. 

\subsubsection{NGC 7662}

NGC 7662 is a triple-shell elliptical, with a bright inner ring,
a fainter outer shell, and a very faint nearly circular halo (Hyung \&
Aller 1997).  This PN also has a large number of complex microstructures,
recently examined using HST imaging by Balick et al. (1998).  They suggest
that there is a prolate elliptical bubble around the central star
aligned perpendicular to the bright ring. The bright ring  
is interpreted as a torus seen at roughly 30$^\circ$\ inclination.  

The spectrum shown in Figure 9 was taken on the bright ring directly SE of
the central star.  The N end of the slit was near the central star, so 
some stellar continuum is seen in the spectrum, which is otherwise dominated
by \ion{H}{1} and \ion{He}{1} lines.

\subsubsection{IC 351}

This compact PN has a double-lobed structure with a round halo (Hua \&
Grundseth 1986; Aaquist \& Kwok 1990; Manchado et al.\
1996). Feibelman, Hyung, \& Aller (1996) obtained UV and visible light
spectra of IC 351 that show it to be a high excitation nebula, but
without the presence of the usual silicon lines, and suggest that the
silicon atoms could be locked up in grains.

Our spectrum (Figure 9) taken centered on the PN suffers somewhat from an incomplete
subtraction of OH airglow lines, due to the sky frames not being taken
properly for the on-source images. The OH lines show up in emission
mainly in the H and K-band portions of the spectrum.  However, the
major features of \ion{H}{1} and \ion{He}{1} emission lines can be
seen.

\subsubsection{IC 418}

The spectrum of this well-studied young, low-excitation PN was
previously shown to be dominated by lines of \ion{H}{1} and
\ion{He}{1} in the near-IR, with a hot dust continuum (Willner et
al.\ 1979; Zhang \& Kwok 1992; Hodapp et al.\ 1994).  Hora et
al.\ (1993) and Paper I showed broad- and narrow-band near-IR images of
the PN, showing the elliptical, double-lobed structure in the IR.

Three positions in the nebula were observed to determine the spectral
variations across the object.  The positions observed were on the central
star, the peak of the E lobe, and in the E halo region outside of the bright
ring (Figure 10).  In the central position, stellar continuum is visible,
rising towards shorter wavelengths.  The nebular lines are similar in the
central and lobe positions.  The halo emission is almost devoid of lines;
there is some faint \Pab\ present as well as \Brg. This is possibly
reflected from the bright lobes.  The main component of the halo emission
is a weak continuum that rises toward longer wavelengths. 

\subsubsection{IC 2149}

This peculiar PN has a bright core and a roughly bipolar nebula
extending approximately E-W, but does not show the usual \h2\ signature
of bipolar PN.  The two spectra shown in Figure 11 were taken centered
on the bright central star, and on the E lobe.  The core spectrum shows strong
stellar continuum, with the nebular lines superimposed.  The E lobe
emission is primarily from lines of \ion{H}{1} and \ion{He}{1}, in
addition to weak continuum emission which is probably reflected from
the central star.

\subsubsection{IC 3568}

This round PN consists of spherical shells, an inner bright one and a
outer halo.  Balick et al.\ (1987) showed that the structure was
consistent with simple hydrodynamic models of PN that are shaped by
interior stellar winds.  The spectrum taken on the N edge of the PN
is shown in Figure 12.  The predominant features are \ion{H}{1}
and \ion{He}{1} emission lines, and low level continuum emission which is
probably reflected from the central star. 

\subsubsection{IC 4593}

Bohigas \& Olguin (1996) obtained spectroscopy and imaging of this PN
which has two inner shells surrounded by an outer highly excited halo. IC
4593 is unusual in that the condensations outside of the inner region are
located asymmetrically in the SW region. 

Two spectra were taken on this PN, one positioned on the central star, and
the other at a position 3\arcsec\ E (Figure 12).  The core shows bright
stellar continuum, with nebular lines superimposed, and \Pab\ absorption. 
The E spectrum has the typical \ion{H}{1} and \ion{He}{1} emission lines. 
The sky subtraction was not of high quality for this spectrum, which
resulted in OH airglow lines showing up in emission in the H and K
spectral regions. 

\subsubsection{J 320}

Images of J 320 (Balick 1987) show it to have a central region that is
elongated roughly E-W, but the low level flux has a N-S elongation
indicating a shell or streamer extending in this direction.  Three spectra
for J 320 are shown in Figure 13, centered on the core (C), offset
1\arcsec\ N, and offset 1\farcs5 N.  The core has a stronger
contribution from stellar continuum, but otherwise the spectra are
similar.  We therefore detect no spectral differences between the N extension 
and the nebula near the core. 

\subsubsection{M 4--18}

This young, low-excitation PN was recently imaged with HST by Dayal et
al.\ (1997) and shows a toroidal shell surrounding the central star. There
is strong mid-IR emission from warm ($\sim$ 200 K) dust that has a similar
morphology but the position angle of the dust emission maxima is
orthogonal to those shown in H$\alpha$ emission. The spectrum of the core
of this PN shown in Figure 14 has strong stellar continuum, as well as
\ion{H}{1} and \ion{He}{1} line emission.  The slit was positioned N-S
across the compact ring of the PN, so the ionized regions of the nebula
are included in this spectrum. 

\subsection{\ion{H}{1} - line and \h2\ emission}

The emission lines present in the spectra of this PNe group contain those
mentioned in the previous section plus lines of molecular hydrogen.  The
\h2\ lines are strongest in the K band, although in some objects there are
lines visible in the H and J bands as well.  In the objects where more
than one slit position was measured, there is often a large change in the
relative line strength of \ion{H}{1} versus \h2\ emission, indicating that
the emission is being produced in different regions of the nebula.  In
general, the \h2\ emission is more likely to be in the outer regions of the
PNe, whereas the \ion{H}{1} emission lines more closely trace the ionized
regions and has similar morphology to the visible appearance. The PNe in
this group all have some bipolar symmetry in their shape, most being of
the ``butterfly'' morphology characterized by narrow equatorial regions and
large bipolar lobes (e.g., M 2--9, Hb 12).  However, some are classified as
elliptical based on the shape of the brightest components.  Tables 5 and 6
list the line identifications and extracted fluxes for the PNe in this
section. 

\subsubsection{Molecular hydrogen excitation in PNe}

A near-infrared \h2\ emission line spectrum can occur through slow
electric quadrupole vibration-rotation transitions in the ground
electronic state. Because the allowed transitions are such that $T_{\rm
ex} \gtrsim 1000$ K is required to produce a detectable near-IR \h2\ spectrum,
special excitation conditions must exist when the near-IR
spectrum is present.  In Paper I we discussed mechanisms of \h2\
excitation in PNe; see also Kastner et al.\ (1996). If detected in
sufficient number, the observed \h2\ line ratios are an excellent
diagnostic for determining the relative importance of shocks and UV
photons in a photodissociation region (PDR) for the excitation of the \h2\
emission.  Even if the excitation mechanism cannot be determined, the
presence of \h2\ emission is important to understanding the conditions in
PNe and how they evolve through wind interactions and photodissociation. 

An analysis of near-IR \h2\ emission can determine an ortho-to-para
(O/P) ratio, the rotational excitation temperature $T_{ex}(J)$, and
the vibrational excitation temperature $T_{ex}(v)$ of the
molecules. If the rotational and vibrational excitation temperatures
differ, then UV excitation is indicated. This is most readily
determined by comparing the column densities in the upper state
vibration-rotation levels with the upper state energy (in temperature
units; see Hora \& Latter 1994, 1996 for a full discussion; see also
Black \& van Dishoeck 1987; Sternberg \& Dalgarno 1989). When many
\h2\ lines are detected, especially those from highly excited levels
that fall in the J band, a much stronger case can be made for the
importance of UV excitation than does the traditional $v = 2\to 1$
S(1) to \s1 line ratio (e.g.\ Hora \& Latter 1994, 1996).  The
comparison of the column densities to the upper state energy levels
has been done for the PNe with detected \h2\ emission and the results
are summarized in Table 8.  Only the rotational excitation
temperatures are listed.  For collisionally- (shock-) excited spectra,
the rotational and vibrational excitation temperatures are coupled and
the same. For UV excited spectra, the vibrational temperature is the
result of a cascade through levels and not a thermal process. The
rotational levels are easily thermalized by collisions. The observed
O/P ratio is in general rather uncertain, especially if only $v = 1$
lines are detected. We did not attempt to determine the O/P ratio for
objects without sufficient line detections. An observed O/P ratio
lower than 3 indicates that the \h2\ emission is not thermally
excited. A subthermal O/P ratio as determined from near-IR spectra is
not caused by the UV excitation process itself, but is a function of
chemistry and density in the PDR (\eg Hora \& Latter 1996; Black \&
van Dishoeck 1987), and might not be indicative of the ``true'' O/P
abundance ratio of the \h2\ (see Sternberg \& Neufeld 1999). Selected
excitation diagrams for several objects that are discussed below are
shown in Figure 33.

\subsubsection{NGC 40}

The morphological classification of NGC 40 is middle elliptical (Balick
1987), which seems to contradict the previously observed strong correlation
between bipolar morphology and \h2\ detection.  However, if one examines
the low-level emission in the N and S regions of this PN (see Paper I),
one can see material that has broken through and expanded beyond the
elliptical shell defined by the E and W bright lobes.  Mellema (1995) has
found the morphology consistent with models of ``barrel''-shaped PNe, which
have roughly cylindrical emission regions slightly bowed outwards at the
equatorial plane, and less dense polar regions.  Higher resolution and
more sensitive optical imaging has recently been carried out by Meaburn et
al.\ (1996) show gas escaping from the polar regions of the PN, with other
filamentary structure in the outer halo. 

This is the first reported detection of \h2\ in NGC 40.  The spectrum was
taken centered on the W lobe, and the \h2\ lines are relatively weak
compared to the \ion{H}{1} and \ion{He}{1} lines from the ionized gas in
this region.  The \h2\ emission was not detected in the narrowband imaging
surveys of Paper I or Kastner et al.\ (1996), so the molecular emission
must be confined to a region near the bright ionized gas that dominates
the spectrum. 
The data suggest that the \h2\ is
shock-excited. However, insufficient line detections make this result less
than firm. 

\subsubsection{NGC 2440}

NGC 2440 is a bipolar PN with complex morphological and spectral
structure.  In the optical, the nebula is bipolar with the major axis in
roughly the E-W direction for the large outer lobes (Balick 1987; Schwartz
1992).  However, there are two bright lobes near the core that are
positioned along an axis roughly perpendicular to the major axis of the
outer lobes.  There are two fainter knots that are also along a roughly
E-W axis, but not aligned with the outer lobes.  There are filaments and
knots throughout the lobes.  Lopez et al.\ (1998) finds up to three
outflowing bipolar structures in the lobes, and find from their kinematic
study that the inner bright lobes (their lobes ``A'' and ``B'') are the
emission maxima from a radially-expanding toroid viewed nearly in the
plane of the sky. 

In the near-IR, the inner pairs of lobes are also prominent, but the large
E-W lobes are not visible (see Paper I).  Instead, there is a circular
outer halo visible in \h2\ that is not quite centered on the inner lobe
structure.  Also visible are faint \h2\ ``spikes'' that extend from the
center to the circular outer halo, roughly in the equatorial plane of the
large optical E-W lobes (Latter \& Hora 1998). 

The spectra shown in Figures 15 and 16 were taken at three
 different positions in
the PN: on the N lobe (of the innermost bright pair of lobes),
on the fainter E knot, and on a clump of \h2\ emission located on the 
NE edge of the outer circular halo (see Paper I, Figure 4a; it is the clump
visible at the upper left corner of the ``\h2\ sub'' image).  The N lobe
exhibits \ion{H}{1} and \ion{He}{1} lines from the ionized gas in this
region, but also has significant \h2\ emission.  There is also strong
[\ion{Fe}{2}] emission at 1.64 and 1.257 \micron.  The E knot also displays
similar  \ion{H}{1}, \ion{He}{1}, [\ion{Fe}{2}], and \h2\ emission, 
although fainter. 
In contrast to the inner regions, the NE clump spectrum in Figure 16
is dominated by \h2\ emission, with the only \ion{H}{1} lines detected
being \Pab\ and \Brg.  There is strong [\ion{Fe}{2}] emission at 1.64 and
1.257 \micron\ in this region as well.  The excitation analysis for
the three positions observed showed that they are
UV-excited, except for the E knot
position for which there is insufficient data. The low value of the
observed O/P ratio is suggestive of the \h2\ emission arising from a PDR
at this location as well. 

Since the inner region of NGC 2440 is morphologically complex and any line
of sight through the PN is likely to intersect several distinct regions,
it is probably the case that the ionized and molecular zones are not mixed
as the spectra might seem to indicate, but that the slit simply includes
several nebular components, or is looking through a PDR and is sampling
both the molecular and the recently ionized gas. 

\subsubsection{NGC 6720}

NGC 6720 (the ``Ring Nebula'') is probably the best-known PN, and is the
archetype for the ring or elliptical morphology that characterizes the
brightest part of the nebula.  The emission is not consistent with a
uniform prolate shell, however, since the ratio of flux between the edge
and center of the ring is higher than expected from a limb-brightened
shell (Lame \& Pogge 1994).  Balick et al.\ (1992) have suggested that NGC
6720 is actually a bipolar PN viewed along the polar axis, based on
narrow-band imaging and high-resolution spectroscopic observations.  This
view is supported by the presence of \h2\ in the nebula and halo, which
correlates strongly with bipolar morphology.  Guerrero, Manchado, \& Chu
(1997) draw different conclusions, however, based on their chemical
abundance and kinematic study of the nebula.  They argue that the Ring has
a prolate ellipsoid structure, with a halo of remnant red giant wind. 

Our spectra of the Ring (Figure 17) were obtained at two positions, one on
the bright ring directly N of the central star, and the second position
several arcseconds further north, off the bright ring but on a moderately
bright (in \h2) position in the halo.  Both positions show bright \h2\
emission, with the lobe position also showing contributions from emission
lines of \ion{H}{1} and \ion{He}{1} from the ionized gas, as one would
expect based on the visible wavelength and IR images showing the
distribution of the line emission. The ring spectrum is strongly UV
excited, indicating it is the PDR interface to the outer molecular shell. 

\subsubsection{NGC 7026}

The late elliptical PN NGC 7026 has two bright lobes on either side (E-W)
of the central star, with fainter bipolar emission extending roughly N-S
from the core.  Cuesta, Phillips, \& Mampaso (1996) obtained optical
spectra and imaging of this object and found kinematically complex
structure, with several separate outflows at the outer edge of an inner
spherical shell, and suggested that the primary shell may be undergoing
breakup in transition to a more typical bipolar outflow structure. 

Two positions were sampled in NGC 7026, shown in Figure 18, centered on
the E and W bright lobes near the central star.  The lobe spectra are
nearly identical, as one might expect from the symmetry in this PN.  The
\h2\ emission in this PN is fairly weak at these positions. This might be
due to the \h2\ being concentrated in other regions of the PN, and not in
the bright ionized lobes that were sampled by the spectra presented here.
We are unable to determine the excitation mechanism. 

\subsubsection{NGC 7027}

NGC 7027 is one of the most highly studied PN at all wavelengths,
particularly in the infrared because of its brightness and wealth of
spectral features.  Treffers et al.\ (1976) obtained a spectrum for
$\lambda = 0.9 - 2.7$ \micron\ with a beam that included the entire
nebula.  They identified the major near-IR spectral components, including
the first detection of \h2\ lines in a PN, and the first detection of the
unidentified line at 2.29 \micron.  Since then, several near-IR spectra
have been published, including Scrimger et al.\ (1978), Smith, Larson, \&
Fink (1981), Rudy et al.\ (1992), and Kelly \& Latter (1995). 

The spectra shown in Figure 19, one taken centered on the W bright lobe,
and the other at the brightest position in \h2\ of the NW lobe  (see
Paper I). Both show \ion{H}{1} and \h2\ emission; the \h2\ is relatively
stronger in the NW position than in the bright lobe. Narrowband imaging
has shown that the \ion{H}{1} and \ion{He}{1} emission is primarily in the
bright inner ring of the nebula, and the \h2\ emission is in what appears
to be bipolar lobes outside of this shell (Graham et al.\ 1993a,b; Paper
I; Latter et al.\ 1998). It has been argued before based on morphology
that the \h2\ is in a PDR (Graham et al.\ 1993a). Our data clearly
demonstrate this to be the case, with the \h2\ showing a strongly UV
excited spectrum in a relatively high density medium (see Figure 33b). 

\subsubsection{\bd30}

The young PN \bd30 is well-studied in the infrared, 
and is remarkable primarily because of its
large IR emission excess.  It has many similarities to NGC 7027, including
its IR morphology and the presence of \h2\ in the near-IR  and 
unidentified IR (UIR) emission features in the mid-IR spectrum, 
which are usually attributed to
polycyclic aromatic hydrocarbons (PAHs).
Rudy et al.\ (1991) obtained a $\lambda = 0.46 -
1.3$ \micron\ spectrum of \bd30; high-resolution visible and near-IR
images were recently obtained by Harrington et al.\ (1997) and Latter et
al.\ (1998), and ground-based near- and mid-IR images have been presented
by Hora et al.\ (1993), Paper I, and Shupe et al.\ (1998). 

Three positions in \bd30 were sampled in the spectra presented in Figure
20; the emission peak on the N lobe of the ring, the E side of the ring,
and on the \h2\ emission region located approximately 3\arcsec\ E of the
ring.  These spectra show a steady progression of decreasing emission from
the ionized gas and increasing molecular emission as one moves east. As
for NGC 7027, the \h2\ emission in \bd30 
is UV excited (Figure 33a) and defines the
PDR (see also Shupe \etal 1998). 

\subsubsection{Hubble 12}

Hubble 12 (Hb 12) has been notable primarily because it represents one of
the clearest cases known of UV excited near-IR fluorescent \h2\ emission
(Dinerstein et al.\ 1988; Ramsay et al.\ 1993).  Our Hb 12 results from
this survey and our imaging survey were presented in a previous paper
(Hora \& Latter 1996); the spectra are reproduced here for comparison with
the rest of the survey.   

Dinerstein et al.\ had mapped the inner structure and found it to be
elliptical surrounding the central star; our deep \h2\ images showed the
emission to be tracing the edges of a cylindrical shell around the star,
with faint bipolar lobes extending N-S.  We also detected [\ion{Fe}{2}]
line emission at 1.64 \micron\ in a position along the edge of the shell.
 The \h2\ line ratios observed were in excellent
agreement with predictions by theoretical \h2\ fluorescence calculations,
and no significant differences were found between the excitation in the
two positions of the nebula that were sampled (see also Luhman \& Rieke
1996). 

\subsubsection{IC 2003}

IC 2003 is a round, high-excitation PN that has a ring of emission, with a
bright knot on the S edge (Manchado et al.\ 1996; Zhang \& Kwok 1998). 
Feibelman (1997) obtained IUE spectra of this PN that shows a wealth of
nebular and stellar lines.  The IR spectra presented in Figure 22 taken
in the center of the PN shows
that there is little continuum from the nebula; the emission is primarily
from lines of \ion{H}{1} in the J, H, and K bands. There is strong unidentified 
emission at 2.286 \micron\ but none detected at 2.199 \micron.  \h2\ emission
is tentatively detected in the K-band, in the \s1, $v=3\to 2$ S(1), and
$v=1\to 0$ Q(1) lines. Each of the lines are detected at roughly a
2$\sigma$ level.  The line fluxes are not reliable or numerous enough to
allow fitting of the line ratios. 

\subsubsection{IRAS 21282+5050}

The young, carbon-rich PN IRAS 21282+5050 has been
identified as having an 07(f)-[WC11] nucleus (Cohen \& Jones 1987) with
possibly a binary at its center.  Strong $^{12}$CO has been detected in a
clumpy expanding shell (Likkel et al.\ 1988) with elongated emission N-S. 
Shibata et al.\ (1989) believe the elongated emission suggests the
presence of a dust torus in the E-W direction; however, Meixner et al.\
(1993) was evidence for a clumpy, expanding elliptical envelope.  The
elongated structure is also seen in the visible (Kwok et al.\ 1993).  Weak
continuum flux at 2 and 6 cm suggests a young PN just beginning to be
ionized (Likkel et al.\ 1994; Meixner et al.\ 1993).  Kwok et al.\ (1993)
believe there has been a recent sharp drop in luminosity based on the
measured CO/FIR ratio. Weak HCO$^+$ and $^{13}$CO are present (Likkel et
al.\ 1988). 

Two positions were sampled on IRAS 21282+5050, centered on the bright core, and
offset approximately 3\arcsec\ N and 3\arcsec\ W. The spectrum of the offset
position is shown in Figure 22.  The core is dominated
by continuum emission from the central star.  Also present are both
emission lines from the ionized gas, and \h2\ features in the K-band. 
The nebula is compact, about 4\arcsec\ in diameter at K (Paper I).  The slit
therefore samples a slice through the entire nebula, and as a result this
spectrum does not necessarily imply that the molecular and ionized gas is
mixed.  The Lobe spectrum shows primarily lines of \h2\ (with the OH night
sky lines showing up in absorption because of imperfect sky subtraction in
this spectrum).  The lack of emission lines due to \ion{H}{1} and
\ion{He}{1} in the Lobe spectrum indicates that the \h2\ emission is
predominantly in the outer regions of the PN. The data are suggestive of
shock excitation, but this should be considered tentative. 

\subsubsection{M 1--16}

M 1--16 is a PN with a near-IR bright central region and bipolar lobes with
fast winds extending at least 35\arcsec\ from the core (Schwartz et al.\
1992; Aspin et al.\ 1993; Sahai et al.\ 1994).  Several spectra were
obtained in this PN scanning across the central region; the two positions
shown in Figure 23 are on the core position and 1\arcsec\ S of the core. 
Both positions show \h2\ emission; the S position is slightly brighter in
both \h2\ and the ionized nebular lines. Our data reveal that the \h2\ is
UV excited in both regions observed. This had be suggested earlier by
Aspin \etal (1993). The core shows a slight rise towards long wavelengths
indicating emission from warm dust continuum. 

\subsubsection{M 1--92}

M 1--92 (``Minkowski's Footprint'') is a bipolar proto-planetary nebula
similar in near-IR appearance to AFGL 618, and has evidence of highly
collimated outflows along the bipolar axis (Paper I; Trammell \& Goodrich
1996 and references therein). Two positions were sampled in M 1--92, one in
the core and one on the NW bipolar lobe.  There are problems with the sky
background subtraction in both spectra, which are most prominent in the
$\lambda = 1.9 - 2.1$ \micron\ region of the spectrum, but also contribute
to a lower signal to noise ratio (S/N) over the whole dataset. 
Nevertheless, the primary characteristics are apparent.  The core region
is dominated by strong warm dust continuum emission.  There is also weak
\Brg\ and \Pab\ emission, but the other most other \ion{H}{1} and
\ion{He}{1} lines are too weak to be detected.  The lobe position shows
weak \h2\ emission. The emission appears to be shock-excited, but the low
excitation suggested by our data is suggestive of UV excitation. Data of
higher S/N are required to discern the dominant excitation mechanism. 
There might also be \h2\ emission near the core that is being masked by
the strong continuum emission.  In both positions, there also seems to be
CO bandhead emission at $\lambda = 2.3 - 2.5$ \micron\, although the S/N
is not high in these regions. 

\subsubsection{M 2--9}

M 2--9 (the ``Butterfly'') is a highly symmetric bipolar nebula, with lobes
extending from opposite sides of a bright central core, nearly in the
plane of the sky.  Bright knots of emission are visible in the lobes at
the N and S ends.  Our results for M 2--9 from this survey were previously
presented in Hora \& Latter (1994), and some of the spectra are reproduced
in Figures 25 and 26 for comparison.  High-resolution imaging in several
near-IR lines indicated that the lobes had a double-shell structure, with
the inner shell dominated by \ion{H}{1} and \ion{He}{1} line emission from
ionized gas and continuum emission scattered from the central source, and
the outer shell (the ``Lobe O'' spectrum in Figure 25) 
of the lobes showing strong \h2\ emission which exhibit a
spectrum consistent with UV excitation in a PDR.  The core region shows a
strong dust continuum component, as well as emission lines of
\ion{H}{1}, \ion{He}{1}, \ion{Fe}{2}, [\ion{Fe}{2}] and \ion{O}{1}.  The N
knot has strong [\ion{Fe}{2}] emission, with relatively weaker \ion{H}{1},
\ion{He}{1}, and \h2\ emission. 

\subsubsection{Vy 2--2}

Vy 2--2 is a compact PN, so very little is known about its morphology. The
spectrum obtained in this survey was taken centered on the bright core and
the slit sampled most or all of the emission from this object.  The
spectrum contains stellar continuum, lines of \ion{H}{1} and \ion{He}{1}
emission from the nebula, and weak \h2\ emission. This detection confirms
the indication of \h2\ emission as reported by Dinerstein et al.\ (1986). 
The spectrum shown in Figure 26 is similar to others in this category,
such as \bd30 and NGC 2440, where several nebular components are
superimposed because of the position and size of the slit. As for those
objects, the \h2\ spectrum in Vy 2--2 is also UV excited. 

\subsection{\h2\ dominated}

The PNe in this group have spectra that primarily contain emission lines
of \h2.  These objects all have bipolar morphology, and most are young or
proto-PNe (PPNe).  The PPNe also have warm continuum dust emission or
stellar continuum that is strongest in the core.  Table 7 lists the line
identifications and fluxes for the PNe in this section. 

\subsubsection{NGC 2346}

NGC 2346 is a PN with faint bipolar lobes seen clearly in \h2\ emission
(\eg Paper I).  The brightest part of the nebula is in the ``equatorial''
region near the central star where the bipolar lobes meet.  Walsh,
Meaburn, \& Whitehead (1991) performed deep imaging and spectroscopy that
showed the full extent of the lobes, and they model the PN as two
ellipsoidal shells that are joined near the central star. The distribution
of the \h2\ emission is similar to the optical (Zuckerman \& Gatley 1988;
Kastner et al.\ 1994; Paper I). 

The near-IR spectrum of NGC 2346 in Figure 27 is dominated by UV-excited
\h2\ emission, as shown in Figure 33d.  The spectrum was obtained with the
slit positioned on the bright condensation to the W of the central star.
There is also weak \Pab\ and \Brg\ emission seen, which is possibly
reflected from near the central star. 

\subsubsection{J 900}

The PN J 900 is a bipolar nebula with an unusual ``jet''-like structure
and an outer shell structure that is seen primarily in \h2\ emission
(Shupe et al.\ 1995; Paper I).  The spectrum of J 900 shown in Figure 27
was obtained at a position N of the brighter lobe just NW of the central
star, centered on the ``jet'' of emission.  Problems with sky-subtraction
caused the J and H-band portions of the spectrum continuum to be slightly
negative.  There is no detected continuum in any part of the spectrum. The
\h2\ spectrum is shock-excited in a moderate velocity wind (Figure 33e). 

\subsubsection{AFGL 618}

AFGL 618 is a carbon-rich, bipolar reflection nebula with a relatively hot
central star ($\approx 30,000$ K), similar spectra in the two lobes, and
the eastern lobe is significantly brighter than the other. In this as in
other ways, the object bears a great resemblance to AFGL 2688 (see below),
despite the fact that their central star temperatures differ by about a
factor of 5.  The visible spectrum shows numerous emission lines
characteristic of
ionized gas (Westbrook et al.\ 1975; Schmidt \& Cohen 1981)
which are scattered by dust into the line of sight, with a small
\ion{H}{2} region surrounding the central object (Carsenty \& Solf 1982;
Kelly, Latter, \& Rieke 1992).  The near-IR spectrum of AFGL 618 is also
dominated by rotation-vibration lines of \h2\ (Thronson 1981, 1983; Latter
\etal 1992; Paper I). 

AFGL 618 exhibits a rich spectrum of molecular line emission (Lo and
Bechis 1976; Knapp et al.\ 1982; Cernicharo et al.\ 1989; Kahane et al.\
1992; Martin-Pintado \& Bachiller 1992; Bachiller et al.\ 1997; Young
1997).  The lines detected include $^{12}$CO, $^{13}$CO, C$^{17}$O,
C$^{18}$O, CS, NH$_3$, HCN, HCO$^+$, CN, and \ion{C}{1}. 

Two of the positions sampled are presented here in Figure 28 -- the core
spectrum and one taken 2\farcs4 E of the core.  Both spectra show strong
\h2\ emission, along with [\ion{Fe}{2}] and weak \Pab\ and \Brg.  In
addition, the core has a warm dust continuum that is apparent throughout
the spectrum, and clear CO bandhead features in the 2.3 -- 2.4 \micron\
region.  The CO features are also present but at lower levels in the
2\farcs4 E spectrum position. Our analysis of the \h2\ spectrum confirms
the earlier results by Latter \etal (1992) -- the spectrum is dominated by
a shock-heated component, but a UV excited component is clearly present as
well (Figure 33c). 

\subsubsection{AFGL 2688}

AFGL 2688 (the ``Egg Nebula'') is a bipolar reflection nebula (Ney et al.\
1975) at visible and near-infrared wavelengths. It has a central star that
exhibits the spectrum of a carbon-rich supergiant (Crampton \etal 1975; Lo
\& Bechis 1976).  Similar in visible appearance to AFGL 915, each lobe
shows two ``jets'' or ``horns'' extending away from the central region
(Crampton et al.\ 1975; Latter \etal 1993; Sahai \etal 1998a).  The lobes
have identical spectra at visible wavelengths, but their brightness
differs significantly (Cohen \& Kuhi 1977).  The near-IR spectrum is
dominated by \h2\ rotation-vibration lines (Thronson 1982; Beckwith 1984;
Latter et al.\ 1993).  A central source is seen in the mid-IR and longer
wavelengths, with fainter extended emission along the axis of the nebula
(Hora et al.\ 1996). There is an enigmatic equatorial region seen in \h2\
emission and might be traced by other molecular species, such as HCN
(Latter \etal 1993; Bieging \& Ngyuen-Quang-Rieu 1996; Sahai \etal 1998b). 

Similar to AFGL 618, this object also has a rich molecular content.
SiC$_2$ is seen in absorption (Cohen \& Kuhi 1977); this feature is
usually found in stars of the highest carbon abundance.  Strong absorption
features of C$_3$ and emission in C$_2$ (Crampton et al.\ 1975) are
present, while C$_2$ is also seen in absorption in reflected light from
the lobes (Bakker et al.\ 1997). The CO $J = 1 \to 0$ line shows three
distinct velocity structures (Kawabe et al.\ 1987; Young et al.\ 1992). 

Our results for this object from this survey were previously presented in
Hora \& Latter (1994, 1995) and our narrowband imaging in Latter et al.\
(1993).  The spectra are reproduced here for comparison with the rest of
the survey.  Spectra were obtained at several positions in the nebula,
including positions along the N lobe, and in the equatorial region (see
Hora \& Latter 1994 for details).  The emission is segregated; the core is
dominated by continuum emission, there are emission lines of C$_2$ and
CN further from the core along the lobes, and the \h2\ emission is
confined to the ends of the lobes and in the equatorial region in what
appears to be a ring or toroidal structure (Latter \etal 1993; Sahai \etal
1998b). Our analysis of the \h2\ line ratios showed that the emission is
collisionally excited in shocks, with no discernible difference between
the emission in the lobes and the equatorial region. 

\subsection{Continuum - dominated}

These are young PNe or PPNe that have strong warm dust continuum and
little line emission.  The strongest component is in general the core,
with most of the emission from an unresolved point source.  In some of the
nebulae, emission structure extends a few arcseconds from the core region. 
Also, in objects such as AFGL 915, they are associated with larger optical
nebulae that extend arcminutes from the core.  In this survey, only the
regions near the core were sampled. 

\subsubsection{AFGL 915}

AFGL 915 (the ``Red Rectangle'') is a carbon-rich biconical reflection
nebula with a metal-depleted spectroscopic binary at its center (Cohen et
al.\ 1975).  The nebula appears axially symmetric and shows spikes running
tangent to the edge of the bicone.  Surrounding the post-AGB star at its
center is a circumbinary disk viewed edge-on (Jura, Balm, \& Kahane 1995)
which could be oxygen-rich (Waters et al.\ 1998). 

C$_2$ and CN are not detected near the binary, though C$_2$ is present in
emission in the reflection lobes.  CH$^+$ (0,0) and (1,0) are detected in
emission (Bakker et al.\ 1997; Balm \& Jura 1992).  CO is underabundant,
with relatively weak emission and broad wings detected (Dayal \& Bieging
1996; Greaves \& Holland 1997; Loup et al.\ 1993; Bujarrabal et al.\
1992).  Glinski et al.\ (1997) found CO and \ion{C}{1} in the UV in both
absorption and emission.  They expect strong CO overtone emission in the
IR based on their observations of hot CO emission and absorption in the
UV.  The object shows ERE (extended red emission) from about $\lambda =
5400$ to 7200 \AA\ and a set of emission bands around 5800 \AA\ (Schmidt,
Cohen, \& Margon 1980) whose carriers might be the same material as the
carriers of the DIBs (diffuse interstellar bands).  This object also shows
strong emission in the PAH bands at 3.3, 7.7, and 11.2 \micron\ (Cohen et
al.\ 1975), which are located predominantly in the lobes and spikes of
emission (Bregman et al.\ 1993; Hora et al.\ 1996). 

Spectra taken at two different positions are shown, one centered on the
core, and the other at 4\arcsec\ S of the core.  Both show strong warm
dust continuum, and the core also has strong CO bandhead emission features
in the $\lambda = 2.3 - 2.4$ \micron\ range. 

\subsubsection{\irc10420}

\irc10420 is a highly evolved, OH/IR star that is thought to be in a
post-red supergiant phase (Jones et al.\ 1993).  The central star seems to
have changed spectral type, transitioning recently to an early A type
(Oudmaijer 1998). Oudmaijer et al. (1996) detected several of the hydrogen
lines in absorption and emission in the near-IR, and Oudmaijer (1998)
presented a high-resolution 0.38 -- 1 \micron\ spectrum showing a large
number of emission and absorption lines.  Recent HST imaging by Humphreys
et al.\ (1997) shows that the circumstellar environment around this star
is extremely complex, with spherical outer shells that extend to a
diameter of 6 arcseconds, and several inner condensations. In the near-
and mid-infrared, bipolar lobes are visible that extend $\sim 2$
arcseconds from the core. 

The spectrum of \irc10420 shown in Figure 31 was taken centered on the
object.  The slit length includes the inner few arcseconds of the object,
although it is dominated by the bright core.  The observed spectrum shows
a bright and relatively featureless continuum. Some \ion{H}{1} lines, e.g.,
\Pab, are seen in absorption.

\subsubsection{M 2--56}

The PPN M 2--56 is a bipolar nebula with a bright central core.  It is
similar in morphology to AFGL 618, although it seems to be at an earlier
evolutionary stage since it does not appear to have an \ion{H}{2} region
(Trammell, Dinerstein, \& Goodrich 1993; Goodrich 1991).  The spectrum of
M 2--56 shown in Figure 31 was taken centered on the core of this PPN.  The
dominant feature is a hot dust continuum that is most prominent in the K
band region of the spectrum.  There are some residual features in the
spectrum from imperfect sky subtraction, mostly in the K band. 

\section{Discussion and Summary} 
\subsection{Spectral Categories} 

The PNe spectra presented in this paper were grouped according to spectral
characteristics as described above.  The groups are an efficient way to
present the data, but also can be seen to correlate strongly with other
characteristics of the PN. 

\subsubsection{Morphology}

The group of \ion{H}{1} - line dominated PNe is composed of primarily
elliptical or round PNe, along with the peculiar or irregular nebulae of
the sample.  In general these PNe are well-known from optical studies,
identified either by their morphology or their optical spectra.  Many of
the PNe in this group of the sample, however, do have IR ``excess''
continuum emission from warm dust, which in some cases prompted their
inclusion in this sample. 

The spectral groups with molecular and/or dust continuum emission are
primarily bipolar.  This classification includes objects such as NGC 6720
which have a ring morphology but are thought to be bipolar viewed pole-on;
Hb 12 which is brightest in \h2\ in the equatorial region and along the
outer edges of the lobes; M 2--9 which is brightest in \h2\ at the edges of
the lobes with no equatorial emission other than at the core; and AFGL
2688 which is brightest along the axis of the bipolar lobes, with \h2\
emission in the equatorial plane.  Clearly this is a heterogeneous group
with a wide range of emission and morphological differences that imply a
range of evolutionary tracks and states. 

\subsubsection{The Carbon-to-Oxygen Ratio}

Carbon stars, although a small fraction of all AGB stars, return about
half of the total mass injected into the ISM by all AGB stars, since
they have on average much higher mass loss rates ($>10^{-4}\ {\rm
M}_{\sun}\ {\rm yr}^{-1}$) than do O-rich objects. The
carbon-to-oxygen (C/O) abundance ratio in PNe has previously been
shown to correlate with morphology (Zuckerman \& Aller 1986), with
bipolar PNe tending to be carbon-rich. It is therefore expected that
the C/O ratio also correlates with the spectral classifications
presented here.  This is in general the case, with the \ion{H}{1} -
line dominated PNe having C/O ratios less than or about 1, whereas the
remaining categories which are dominated by the bipolar PNe have C/O
ratios $>$ 1, as reported by Zuckerman \& Aller (1986) and Rola \&
Stasi\'nska (1994).  Rola \& Stasi\'nska discuss problems with
previous determinations of the C/O ratio, and use different criteria
that result in a slightly lower percentage of carbon-rich PNe (35\%)
than others.  Their ratios are used in the discussion below.

The morphology of PN also has been shown to depend on the progenitor mass
(see Corradi \& Schwarz 1995), with the bipolar PN being more massive than
other morphological types.  This relationship, along with the link between
carbon abundance and morphology, suggests that carbon stars are the
progenitors of bipolar PN and those with a large amount of molecular
material.  The mechanisms that cause massive carbon-rich stars to
preferentially form bipolar PN are still not understood. 

There are some exceptions to the correlation of morphological type to C/O
ratio; in particular, NGC 6543 has a much higher value (9.55) than the
others in the class. In the other extreme, NGC 2346 stands out as having a
low C/O ratio (0.35) compared to other bipolar PNe in the \h2\ - dominated
group.  This object is a much more evolved object than the others in its
group (e.g., AFGL 2688), and exhibits weak \Pab\ emission, showing that
the ionized gas is present although weak relative to the molecular
emission in the nebula. 

\subsection{Spectral Sampling of Morphological Features}

This survey has differed from many previous investigations in that a
short, narrow slit was used to obtain the data, rather than a large beam
that could include most or all of the nebula.  Because of this, one cannot
easily use the spectra presented here to model the PNe in a global sense,
if that requires a measurement of the total flux from the object.  Also,
if a complete census of emission lines were required, some might be missed
if there were variations of emission characteristics across the nebula and
certain regions were not sampled. 

The spatial selectivity that prevents viewing the entire PN at once,
however, has proven to be an advantage when trying to examine various
aspects of the PN, including variations across the nebula, as a function
of distance from the central star, or in examining certain morphological
features.  For example, in M 2--9 and NGC 2440, the emission of the lobe
walls and emission knots were separately sampled, which showed the large
spectral differences in these regions.  This information is important for
modeling the structure and formation of the PN.  Another reason why the
small aperture is useful is that if the emission from a group of lines
such as \h2\ is to be modeled, it is important to compare the emission
from a clump of material where the conditions do not vary greatly over its
size.  For example, the emission from \h2\ present very close to the
central star in a strong UV field could be quite different from \h2\
emission from the outer parts of the halo.  Also, the small slit has aided
in detecting weak \h2\ emission from several PNe such as NGC 40, where
detection would have been difficult if the central star and the rest of
the nebula could not be excluded from the measurement. 

\subsection{Summary of Molecular Hydrogen Emission in PNe}
 
A long-standing problem in the interpretation of \h2\ emission from
interstellar and circumstellar environments is understanding the
excitation mechanism. Three fundamental mechanisms are possible.  One is
excitation of a near-infrared fluorescence spectrum resulting from a
rotational-vibrational cascade in the ground electronic state following
electronic excitation by the absorption a UV photon in the Lyman and
Werner bands (Black \& van Dishoeck 1987). A second excitation mechanism
is collisional excitation in a warm gas ($T_{\rm K} \gtrsim 1600$ K).
While UV excitation in a low density gas produces an easily identifiable
spectrum, the level populations can be driven to produce thermal line
ratios when the UV flux is large and densities begin to exceed \about\
$10^4$ \cm3 (Sternberg \& Dalgarno 1989).  Detailed spectral and
morphological analysis are often required to determine an origin of the
near-IR spectrum. A third excitation mechanism is formation of \h2\ on
the surfaces of dust grains and in the gas phase. While potentially
important in isolated regions of certain objects, we do not consider this
to be generally important in PNe and PPNe relative to the other two
processes. This is because molecular formation in PNe is relatively slow
compared to dissociation rates. 

In PNe and PPNe, the situation can be complicated by both dominant
excitation mechanisms being present simultaneously, and in different
forms.  Several ways of exciting near-IR \h2\ emission have been
identified as possible: direct thermal excitation in warm gas created
behind moderate velocity shocks, direct excitation by UV photons from
the hot central star, somewhat indirectly by collisional excitation in
warm gas created by rapid grain streaming (e.g. Jura \& Kroto 1990), and
excitation through absorption of \Lya photons (by an accidental
resonances with the B$^1\Sigma_u^+ - {\rm X}^1\Sigma_g^+$ $v = 1 - 2$ P(5)
and R(6) transitions of \h2) which can be generated in a nearby strong
shock (\eg Black \& van Dishoeck 1987). 

The first two mechanisms have been identified in several PNe, such
as thermal excitation in AFGL 2688 (\eg Hora \& Latter 1994; Sahai \etal
1998a), pure UV excitation in a low density gas around Hb 12 (\eg
Dinerstein \etal 1988; Hora \& Latter 1996; Luhman \& Rieke 1996), and UV
excitation in a high density gas in M 2--9 (Hora \& Latter 1994) and NGC
7027 (Graham \etal 1993b; this paper).  A combined spectrum was found from
a detailed analysis of AFGL 618 (Latter \etal 1992; this paper).  While
the form of the excitation might be apparent for these and other objects, it
is not always evident what is the source of the warm gas or UV photons.
Winds are present in AFGL 2688 which could directly heat the gas through shocks,
but considerable grain streaming is likely taking place as well (see Jura
\& Kroto 1990). 

Very fast winds and dissociating shocks are present in AFGL 618, M 2--9
(\eg Kelly, Latter, \& Hora 1998), and M 1--16 (Sahai \etal 1994; Schwarz
1992), and all show clear evidence of UV excitation. In addition, the
photon path to the \h2\ emitting regions is not in a direct
line-of-sight to the central star, which for photons coming from the
central star suggests scattering in what is a fairly low density medium. 
Alternatively, we are seeing in each of these objects excitation of \h2\ at
the bipolar lobe walls by UV photons generated within strong shocks
produced by the fast winds. This hypothesis was explored through detailed
modeling by Latter \etal (1992) of AFGL 618, but the high relative
intensity of the thermally excited emission and poor spatial resolution
limited this analysis. The presence of very fast winds in the lobes of
each of these objects, and the presence of UV excited \h2\ emission at the
lobe walls strongly suggests that indirect excitation of the \h2\ is
occurring by interactions with photons generated by wind-produced
shocks. Detailed modeling of sensitive, high spatial resolution
spectra is required. It is also evident, in general, that without detailed
spectra, \h2\ is a rather poor diagnostic of overall conditions in PNe and
PPNe. 

If we conclude that all of the ways to excite \h2\ in PNe and PPNe listed
above are present and important, what does this imply for our
understanding of these objects and the utility of \h2\ as a diagnostic? It
is now well understood that the presence of molecular emission from PNe
and PPNe is tied to the morphology of the objects such that if molecular
emission is present, the object has a bipolar morphology (\eg Zuckerman \&
Gatley 1988; Latter \etal 1996; Kastner \etal 1996, and references
therein).  We have argued that \h2\ emission is excited in multiple
ways in PNe and PPNe. While special conditions are required for \h2
emission to be seen in near-IR spectra, the conditions that drive the
excitation are common in all PNe and PPNe and are not clearly dependent on
morphological type.  A conclusion that can be drawn from this argument
alone is that molecular material is present in nebulae with a bipolar
morphology and a significant amount of molecular material is $not$ present
in other morphological types. Therefore, objects that have a bipolar
morphology must have a dense, high mass envelope in which the molecular
material can be shielded and survive dissociation for relatively long
times -- suggesting a high mass loss rate and a high mass progenitor star.
A correlation between bipolar morphology and high mass progenitor stars
has been found by others (\eg Corradi \& Schwarz 1995). It is apparent 
that the presence of \h2\ emission in a PN is not tied to directly to the
morphology, but that the bipolar morphology is intimately related to the
density and mass of the circumstellar envelope, and therefore the mass of
the progenitor star. Why high mass, high mass loss rate asymptotic giant
branch stars shed material in an axisymmetric, not spherical, way remains
a mystery. 

\acknowledgments 

We thank Xander Tielens and David Hollenbach for useful discussions and
encouragement. We acknowledge support from NASA grant 399-20-61 from the
Long Term Space Astrophysics Program.  WBL was supported during part of
this study by a National Research Council Research Associateship.

\newpage
%
%  FIGURE 1
%
% 
\begin{figure}
\plotfiddle{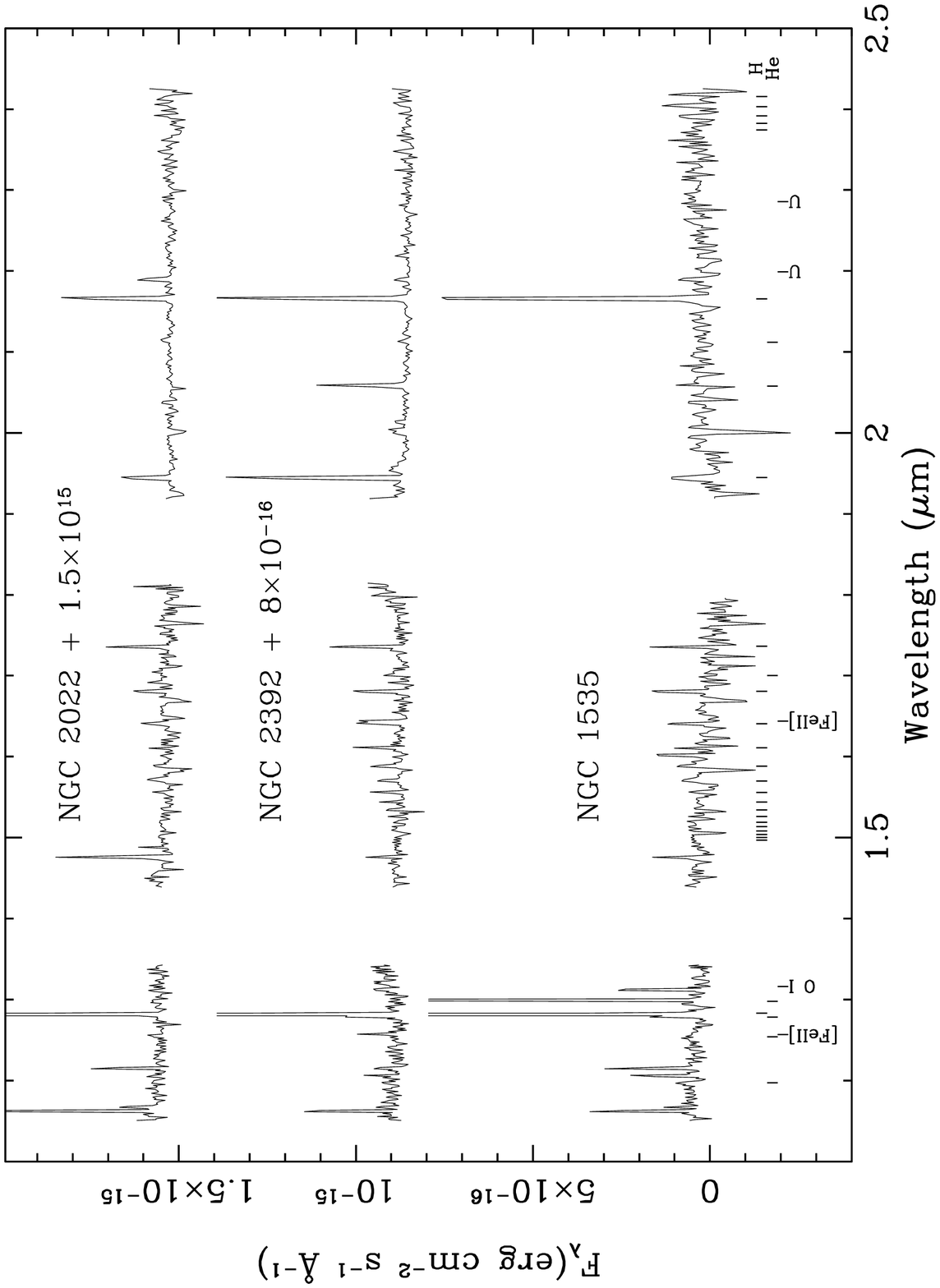}{4in}{270}{58}{58}{-250}{320}
\caption{Spectra of NGC 1535, NGC 2022, and NGC 2392.  The spectra
have been offset and scaled by the constants shown in the plot labels.
The brightest lines in the lower spectra in the figure have been clipped
to keep them from overlapping with the spectra above them.
At the bottom, some of the prominent lines have been labeled with
small vertical lines.  The first row are \ion{H}{1}, the second row
\ion{He}{1}, and below the rows individual lines have been indicated.
These are the same for all plots in this
spectral grouping  for means of comparison; they do not
necessarily indicate that the lines were detected in any or all of the
spectra plotted in the figure.
The spectral data
points between $\lambda = 1.32 - 1.42$ \micron\ and 1.8 -- 1.9 \micron\
are in regions of poor atmospheric transmission and are not plotted.
The positions in the nebula where these spectra were taken is given in
Table 1.}
\end{figure}
%
%  FIGURE 2
%
% 
\begin{figure}
\plotfiddle{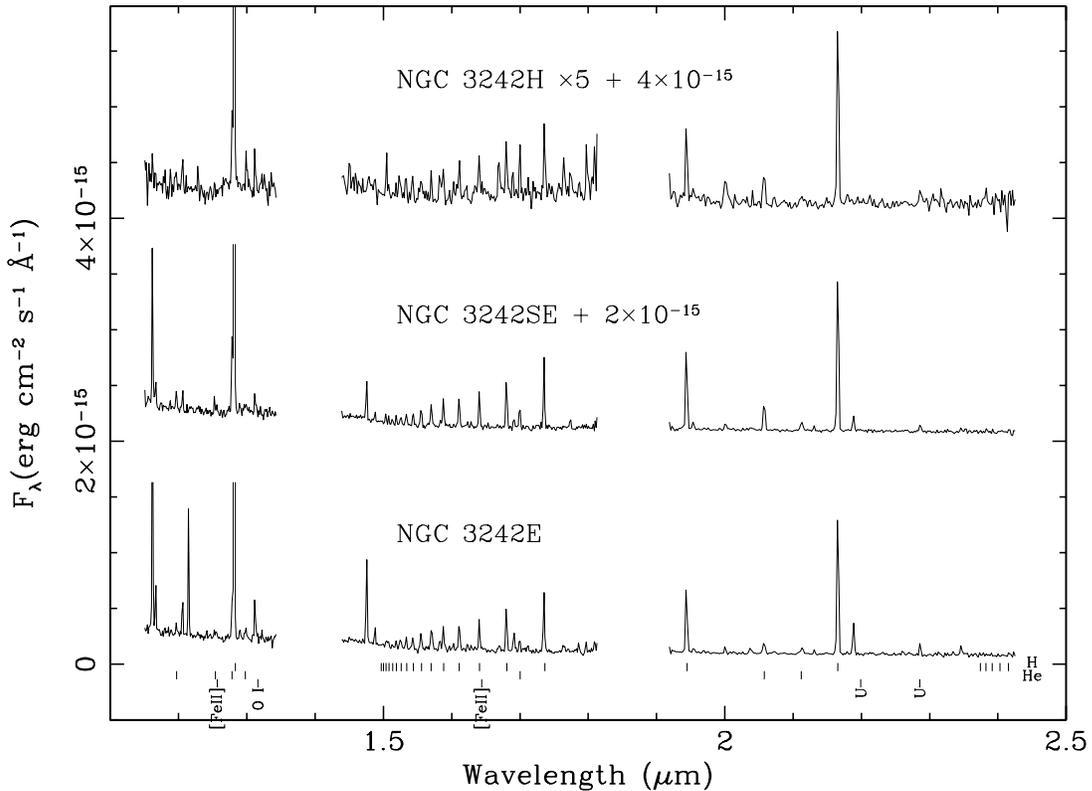}{4in}{270}{58}{58}{-250}{320}
\caption{NGC 3242 (see caption to Figure 1).}
\end{figure}
%
%  FIGURE 3
%
% 
\begin{figure}
\plotfiddle{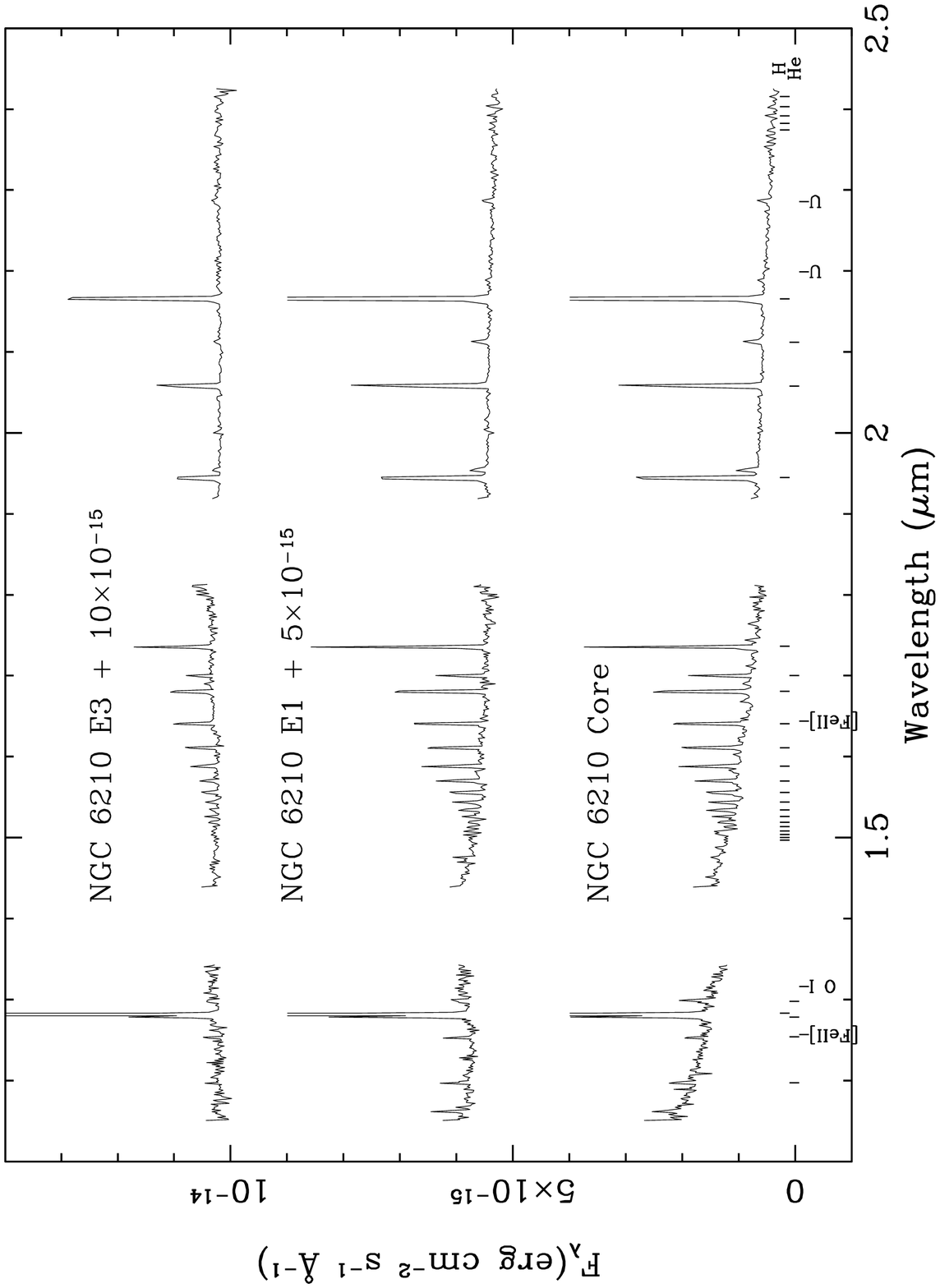}{4in}{270}{58}{58}{-250}{320}
\caption{NGC 6210 (see caption to Figure 1).}
\end{figure}
%
%  FIGURE 4
%
% 
\begin{figure}
\plotfiddle{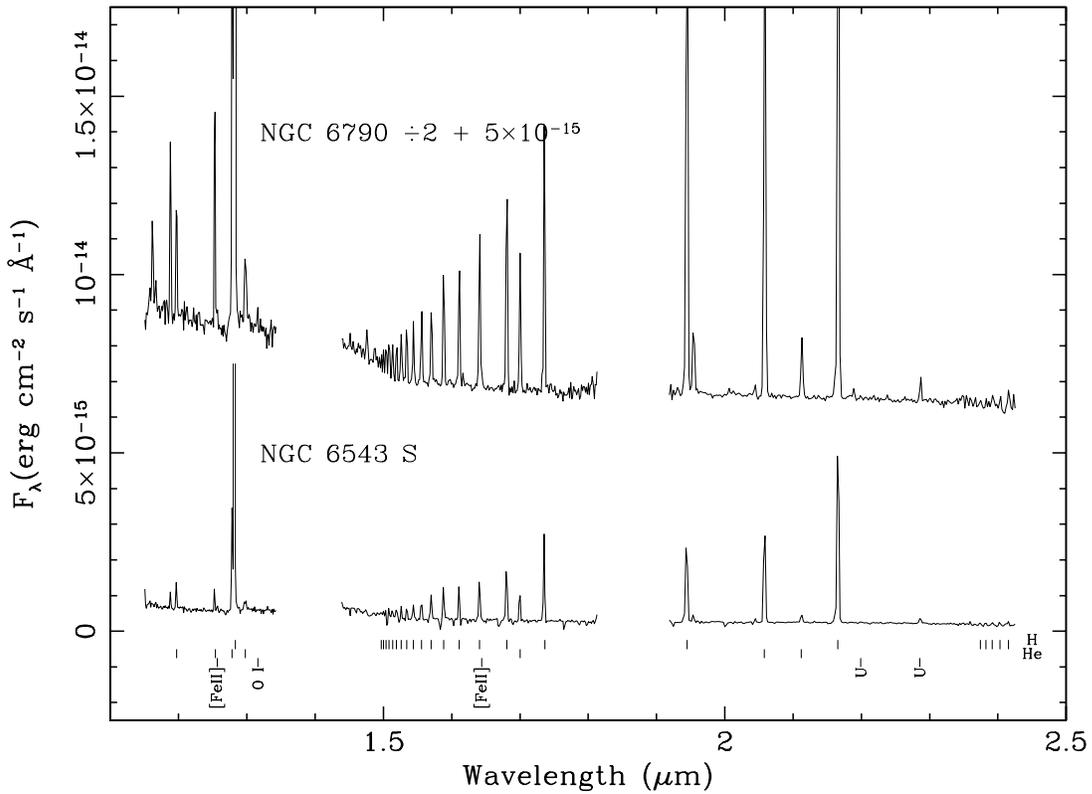}{4in}{270}{58}{58}{-250}{320}
\caption{NGC 6543 and NGC 6790 (see caption to Figure 1).}
\end{figure}
%
%  FIGURE 5
%
% 
\begin{figure}
\plotfiddle{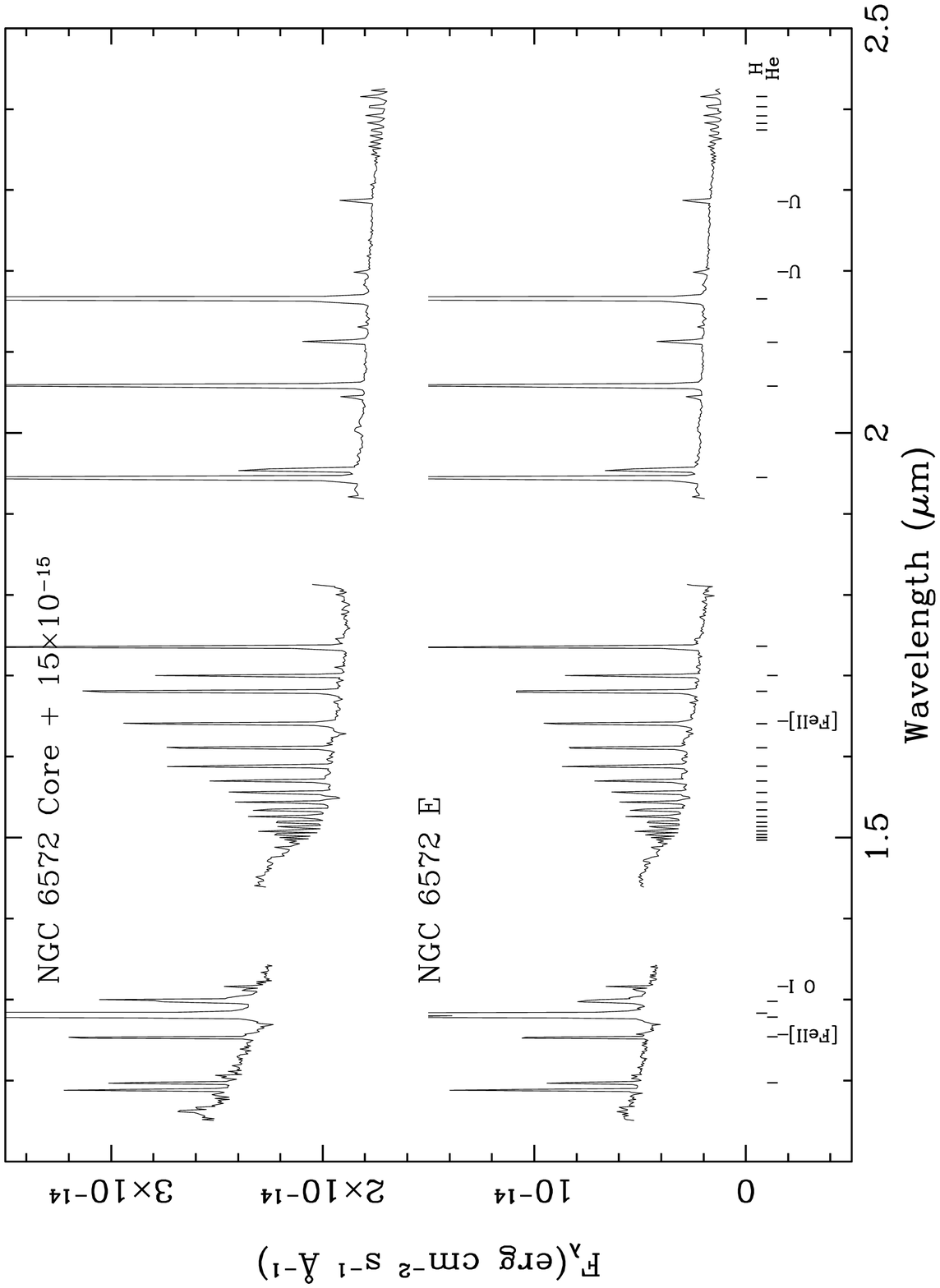}{4in}{270}{58}{58}{-250}{320}
\caption{NGC 6572 (see caption to Figure 1).}
\end{figure}
%
%  FIGURE 6
%
% 
\begin{figure}
\plotfiddle{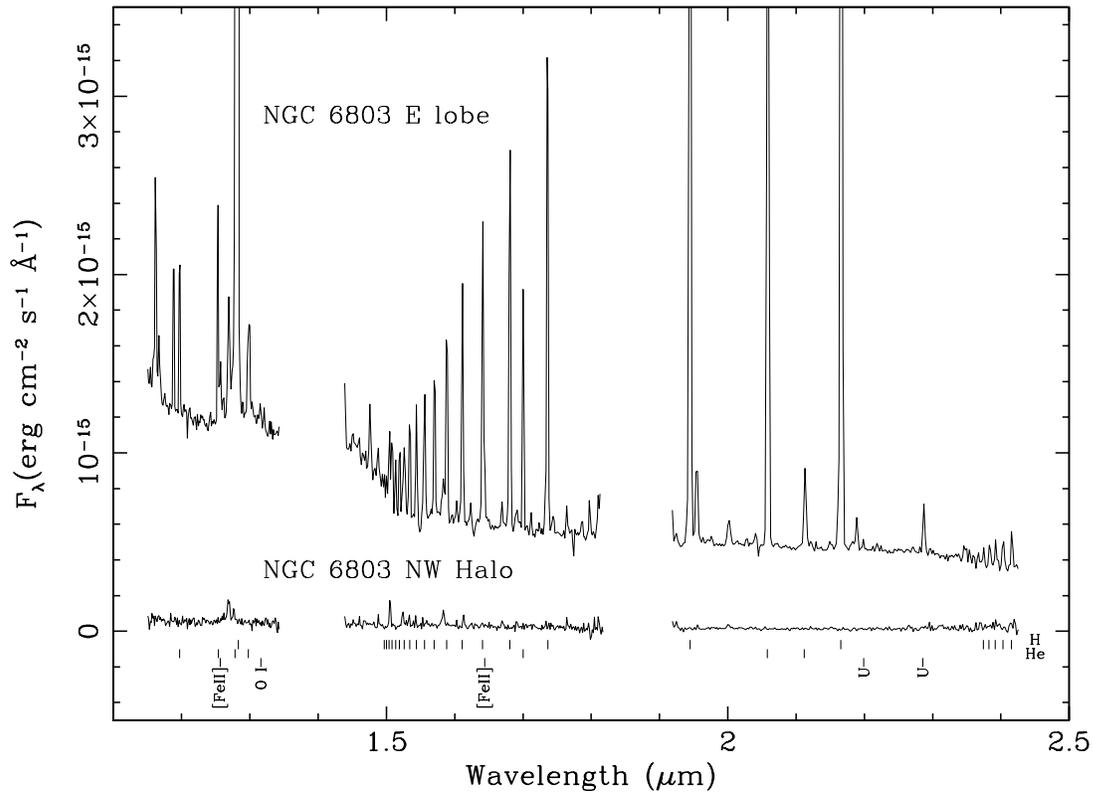}{4in}{270}{58}{58}{-250}{320}
\caption{NGC 6803 (see caption to Figure 1).}
\end{figure}
%
%  FIGURE 7
%
% 
\begin{figure}
\plotfiddle{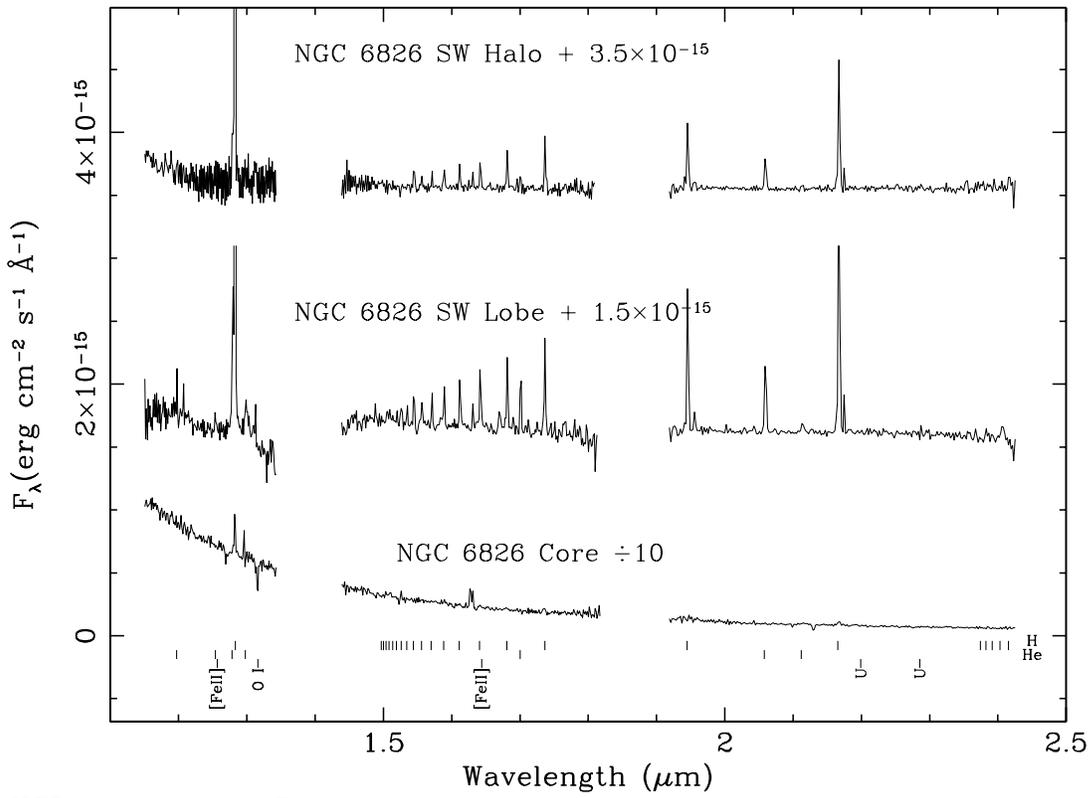}{4in}{270}{58}{58}{-250}{320}
\caption{NGC 6826 (see caption to Figure 1).}
\end{figure}
%
%  FIGURE 8
%
% 
\begin{figure}
\plotfiddle{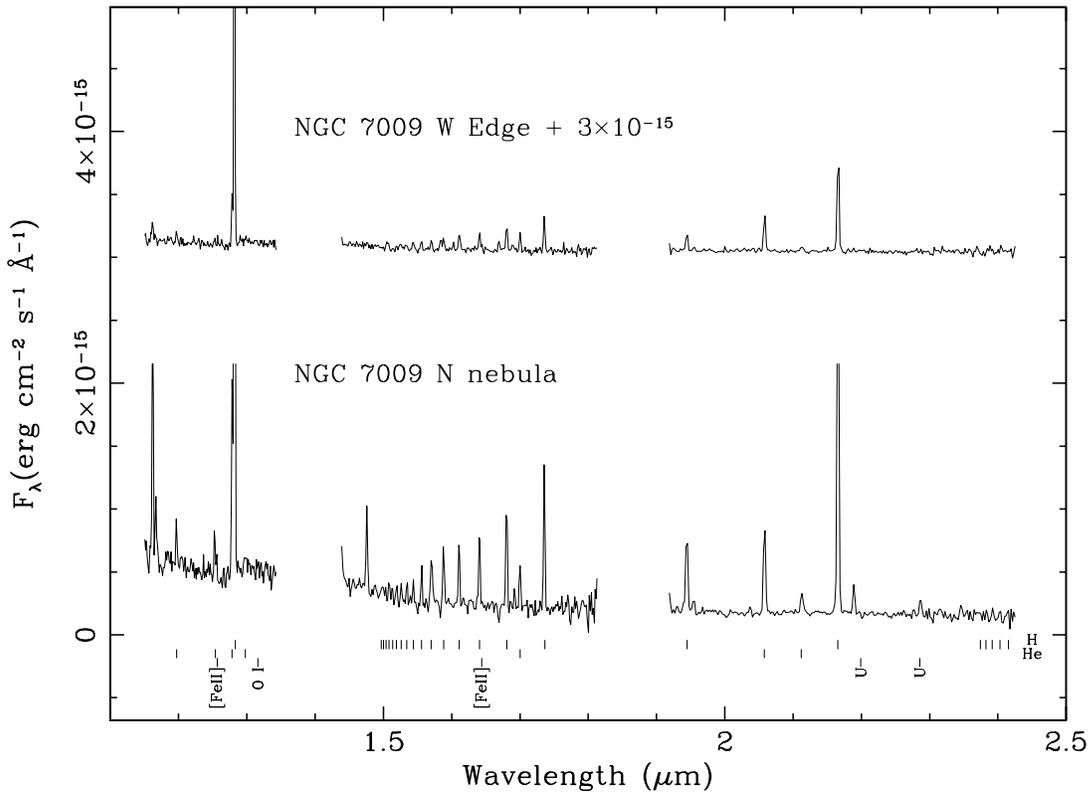}{4in}{270}{58}{58}{-250}{320}
\caption{NGC 7009 (see caption to Figure 1).}
\end{figure}
%
%  FIGURE 8
%
% 
\begin{figure}
\plotfiddle{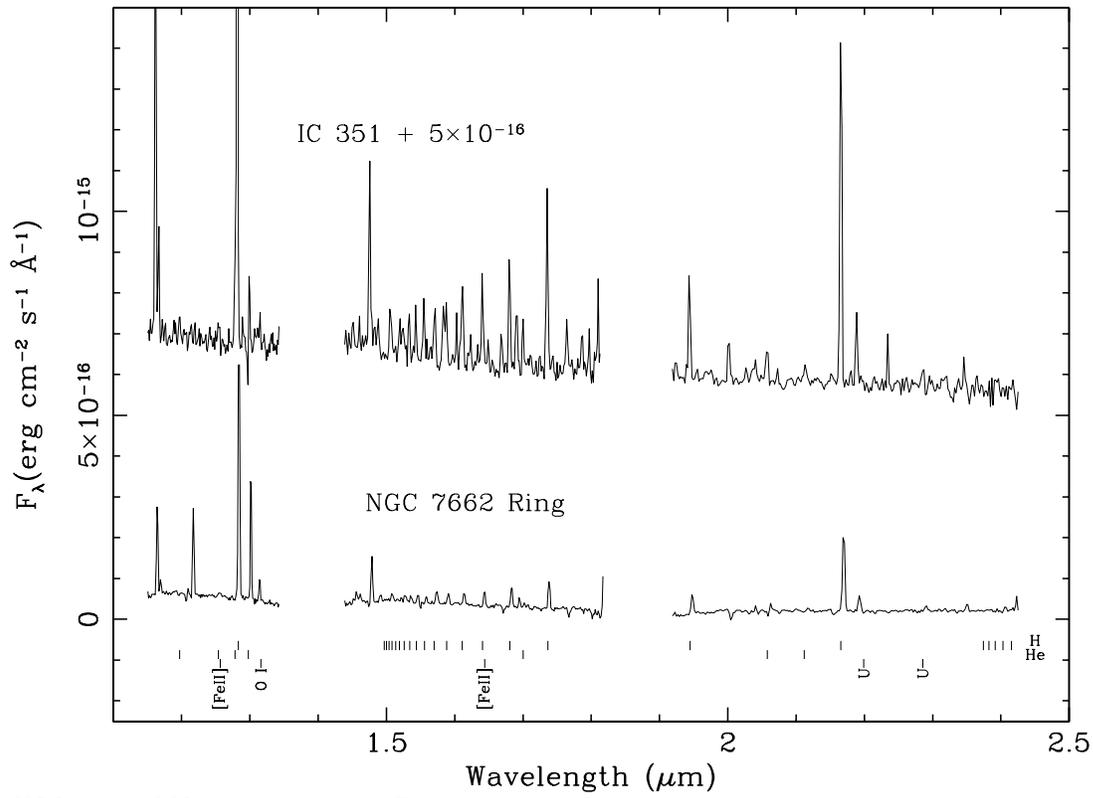}{4in}{270}{58}{58}{-250}{320}
\caption{NGC 7662 and IC 351 (see caption to Figure 1).}
\end{figure}
%
%  FIGURE 10
%
% 
\begin{figure}
\plotfiddle{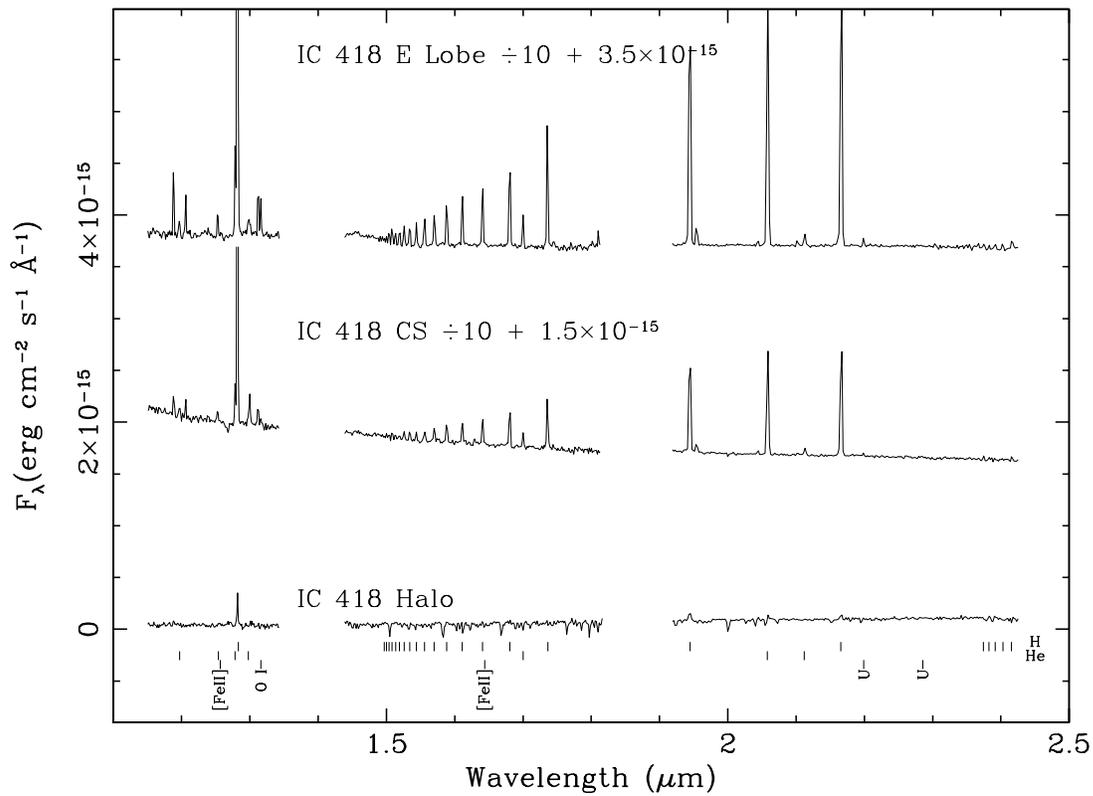}{4in}{270}{58}{58}{-250}{320}
\caption{IC 418 (see caption to Figure 1).}
\end{figure}
%
%  FIGURE 11
%
% 
\begin{figure}
\plotfiddle{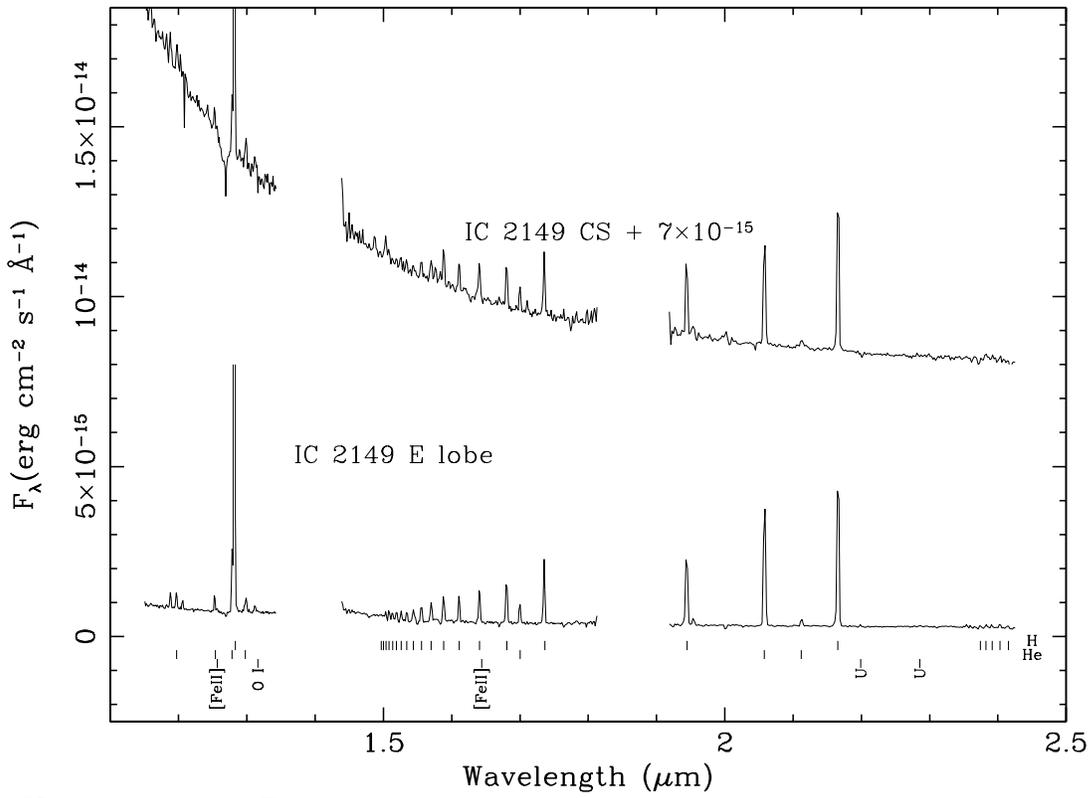}{4in}{270}{58}{58}{-250}{320}
\caption{IC 2149 (see caption to Figure 1).}
\end{figure}
%
%  FIGURE 12
%
% 
\begin{figure}
\plotfiddle{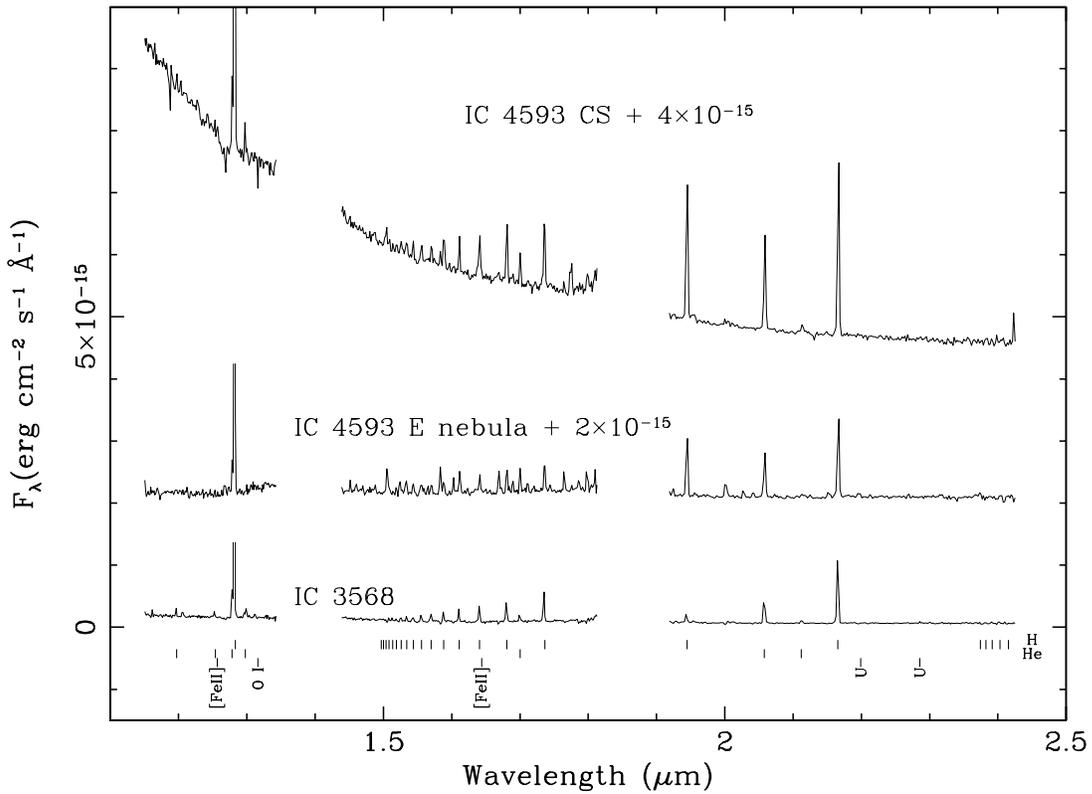}{4in}{270}{58}{58}{-250}{320}
\caption{IC 3568 and IC 4593 (see caption to Figure 1).}
\end{figure}
%
%  FIGURE 13
%
% 
\begin{figure}
\plotfiddle{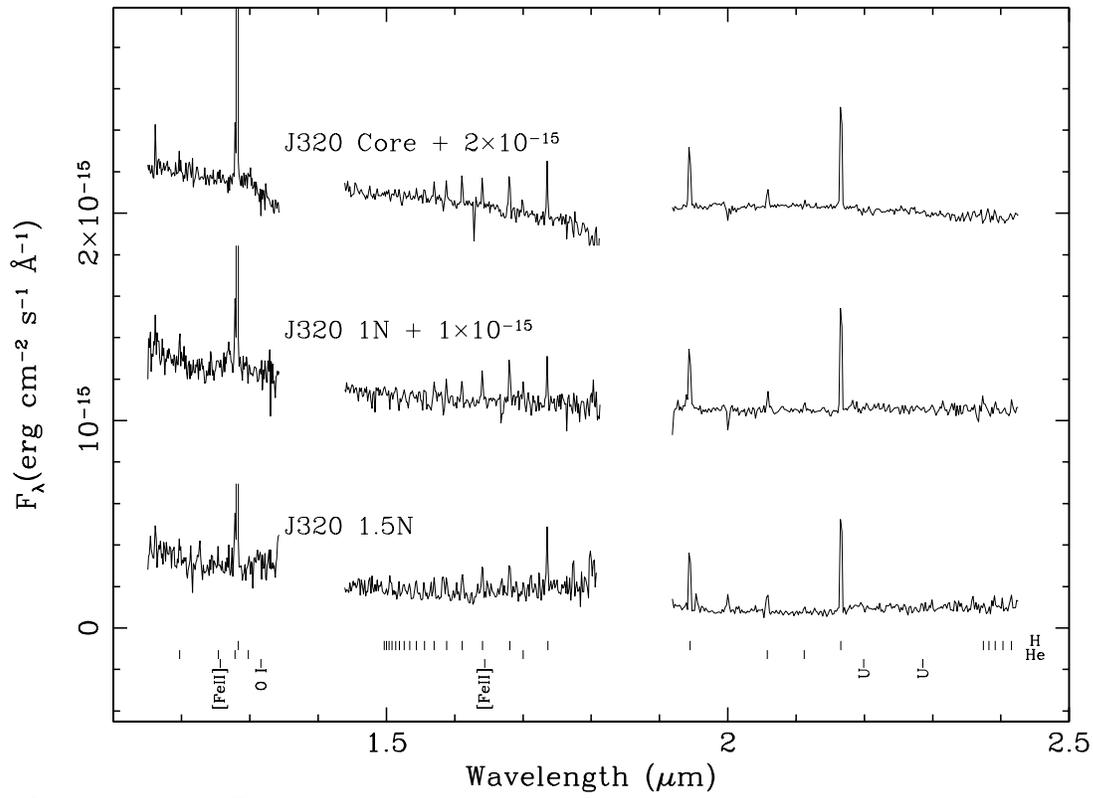}{4in}{270}{58}{58}{-250}{320}
\caption{J320 (see caption to Figure 1).}
\end{figure}
%
%  FIGURE 14
%
% 
\begin{figure}
\plotfiddle{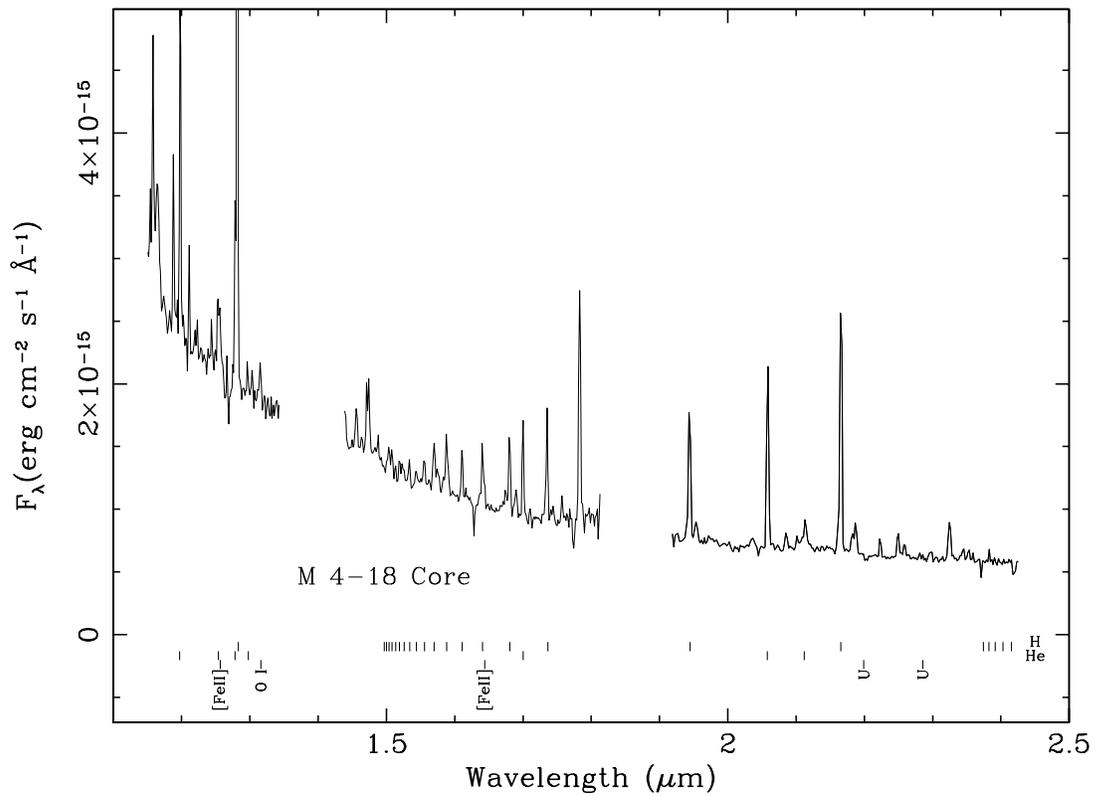}{4in}{270}{58}{58}{-250}{320}
\caption{M 4--18 (see caption to Figure 1).}
\end{figure}
%
%  FIGURE 15
%
% 
\begin{figure}
\plotfiddle{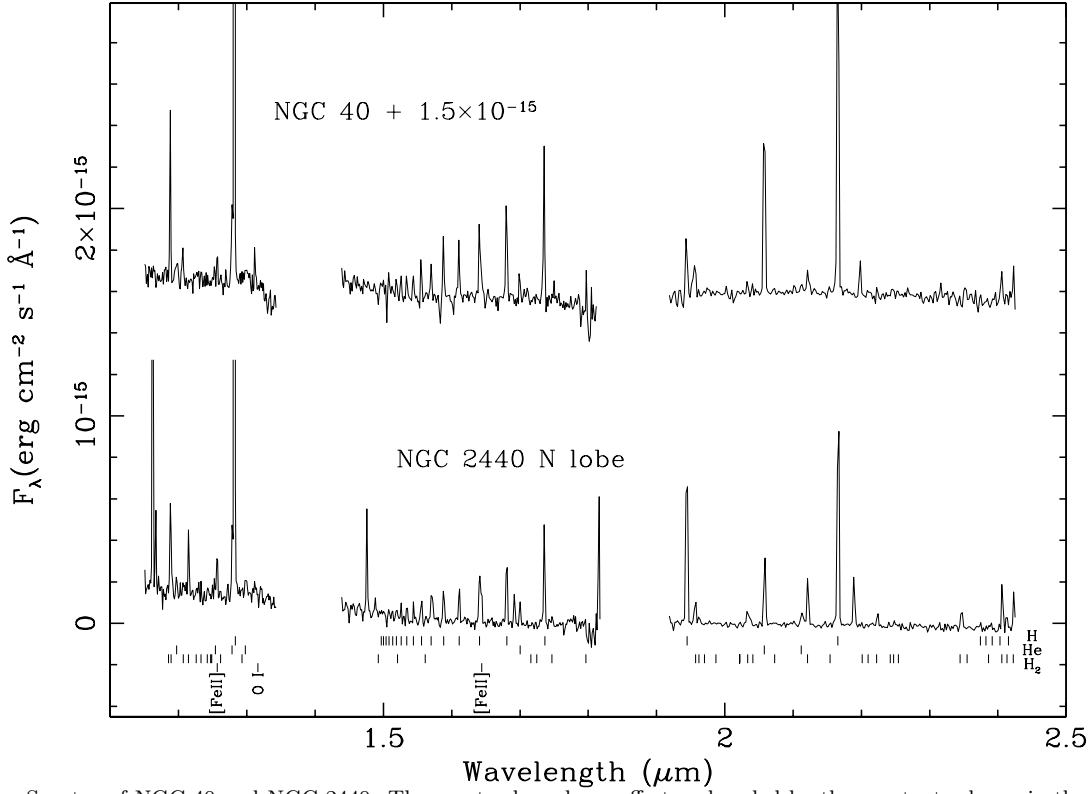}{4in}{270}{58}{58}{-250}{320}
\caption{Spectra of NGC 40 and NGC 2440. The spectra have been offset
and scaled by the constants shown in the plot labels.  At the bottom,
some of the prominent lines have been labeled with small vertical
lines.  The first row are \ion{H}{1}, the second row \ion{He}{1}, and
the third row \h2.  
These are the same for all plots in this
spectral grouping  for means of comparison; they do not
necessarily indicate that the lines were detected in any or all of the
spectra plotted in the figure.
The spectral data
points between $\lambda = 1.32 - 1.42$ \micron\ and 1.8 -- 1.9 \micron\
are in regions of poor atmospheric transmission and are not plotted.
Below the rows individual lines have been
indicated.  The positions in the nebula where these spectra were taken
is given in Table 1.}
\end{figure}
%
%  FIGURE 16
%
% 
\begin{figure}
\plotfiddle{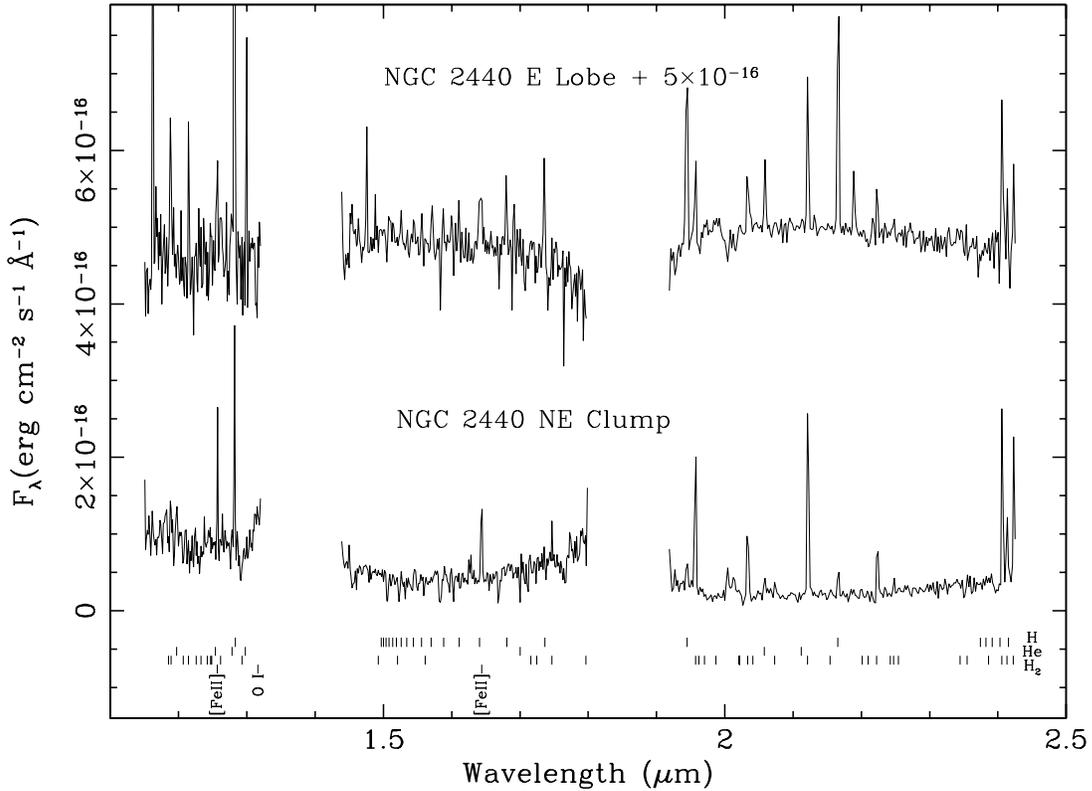}{4in}{270}{58}{58}{-250}{320}
\caption{NGC 2440 (see caption to Figure 15).}
\end{figure}

%
%
%  FIGURE 17
%
% 
\begin{figure}
\plotfiddle{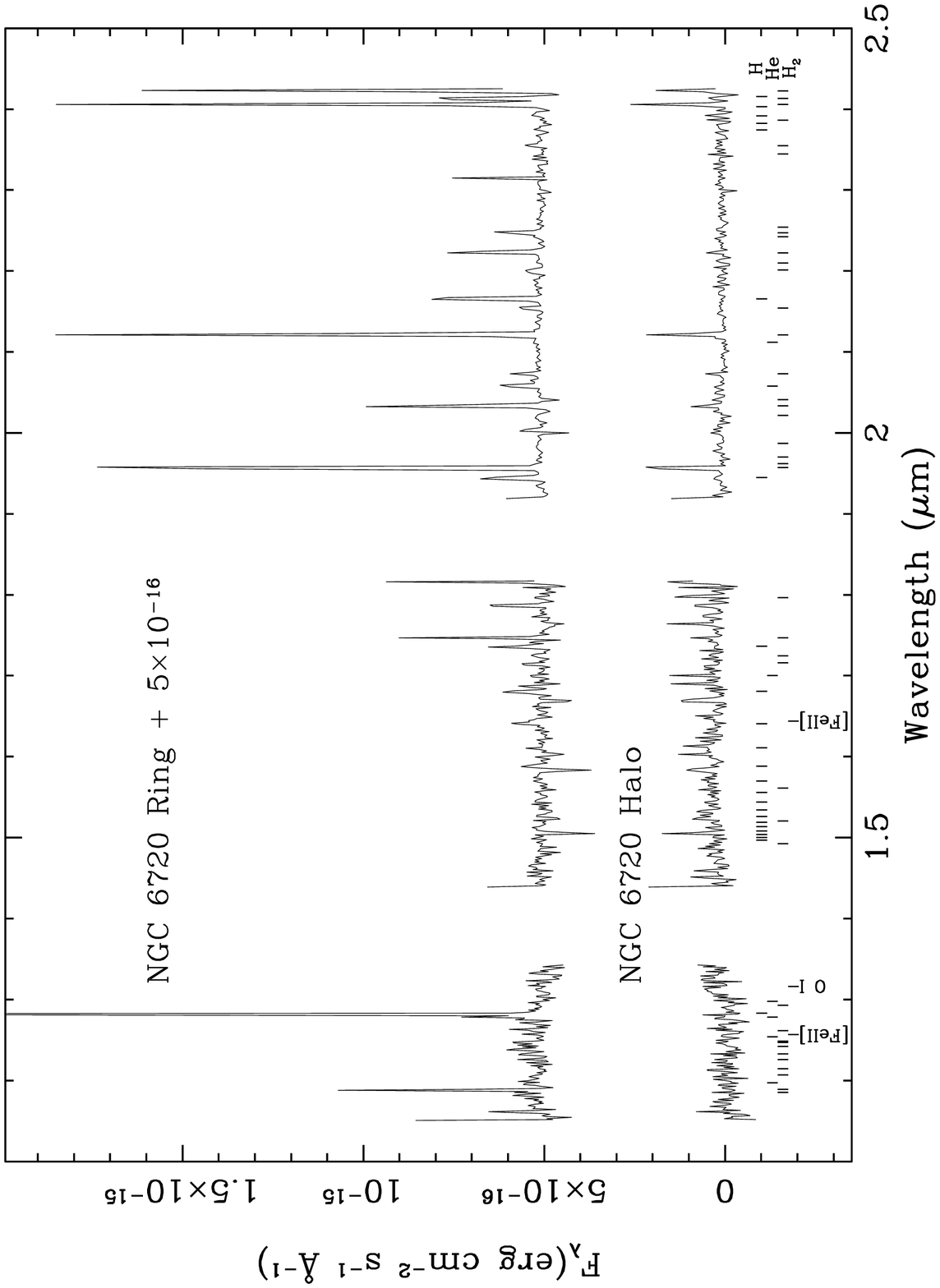}{4in}{270}{58}{58}{-250}{320}
\caption{NGC 6720 (see caption to Figure 15).}
\end{figure}
%
%  FIGURE 18
%
% 
\begin{figure}
\plotfiddle{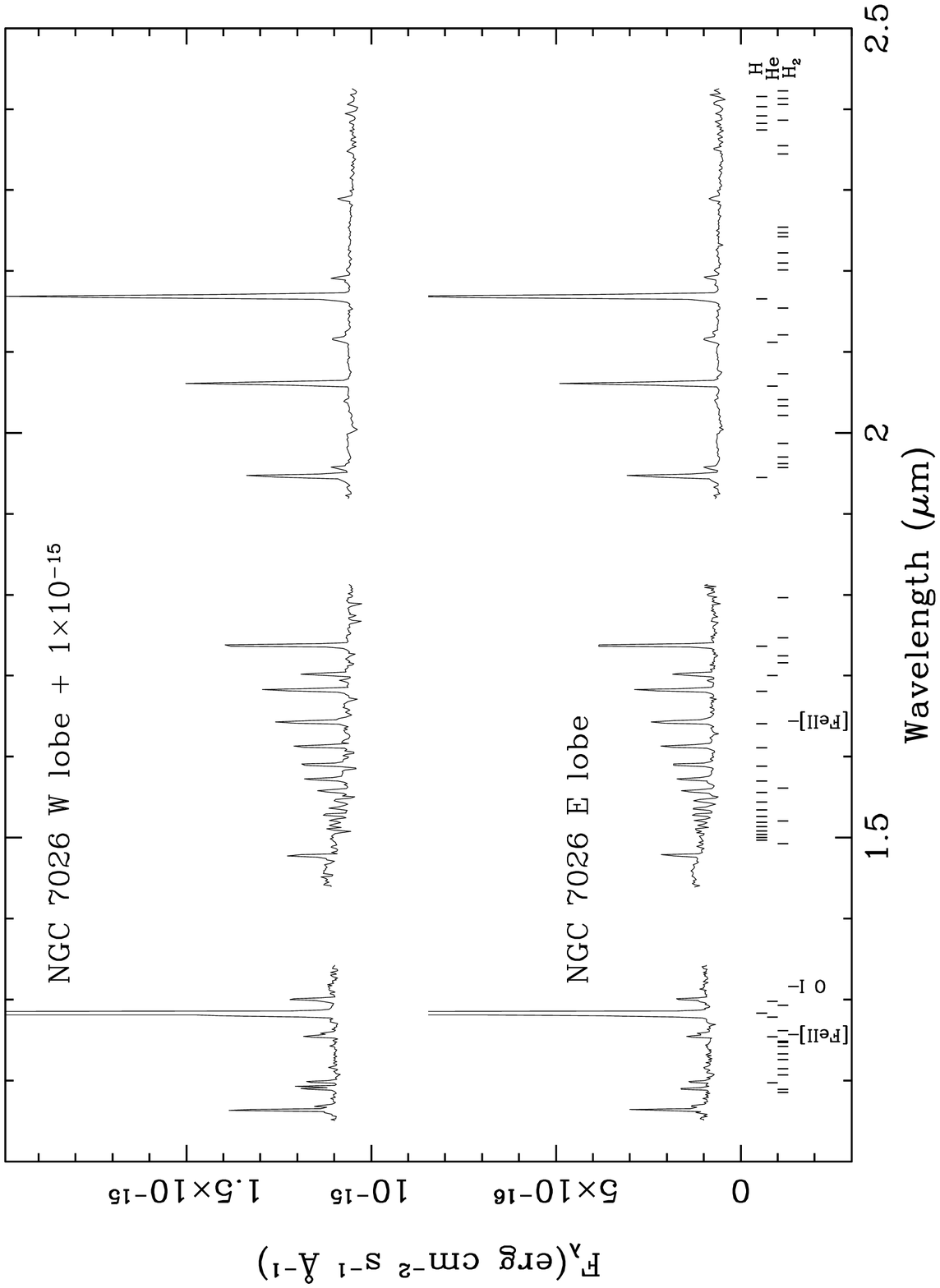}{4in}{270}{58}{58}{-250}{320}
\caption{NGC 7026 (see caption to Figure 15).}
\end{figure}
\clearpage
%
%  FIGURE 19
%
% 
\begin{figure}
\plotfiddle{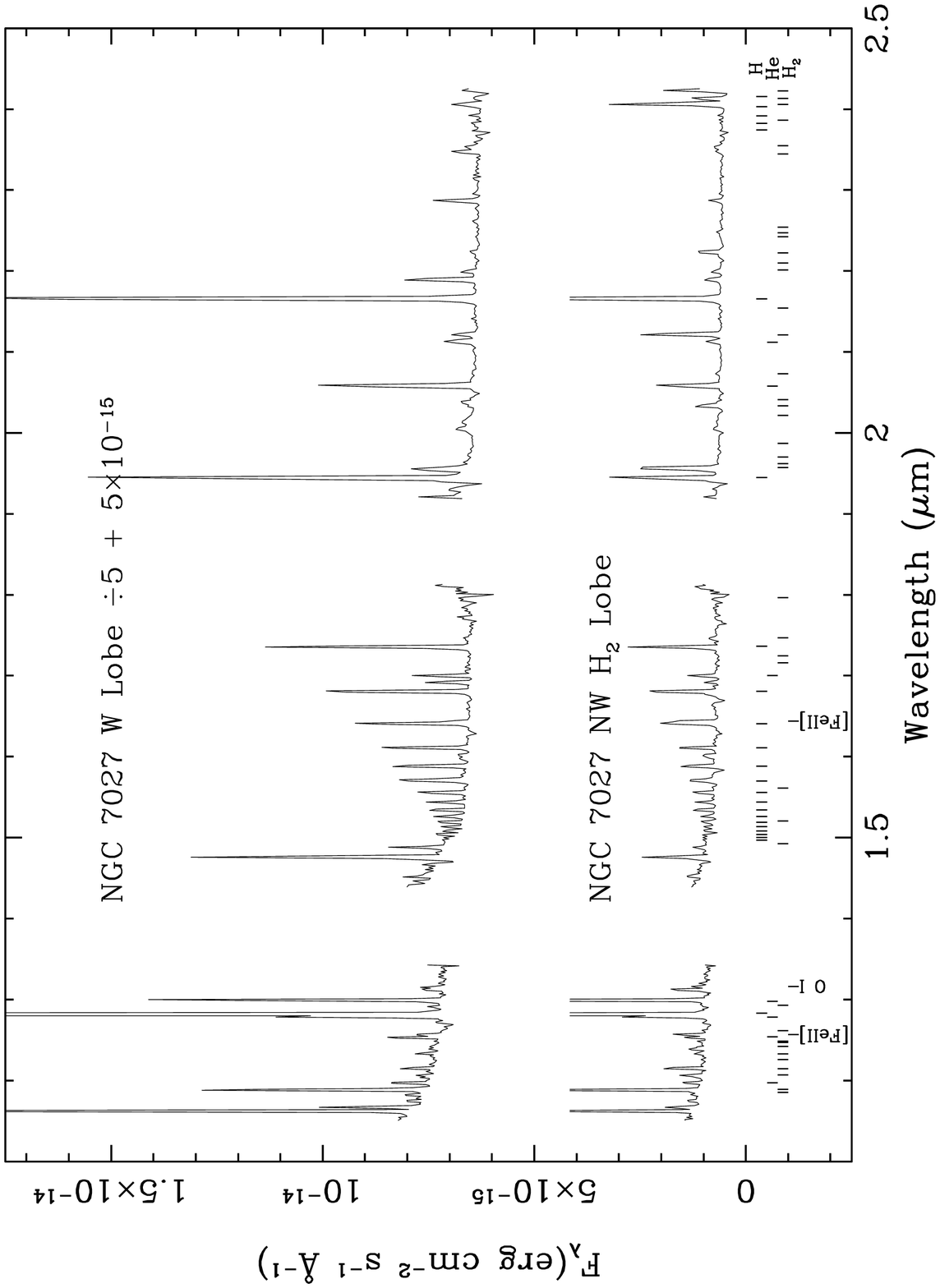}{4in}{270}{58}{58}{-250}{320}
\caption{NGC 7027 (see caption to Figure 15).}
\end{figure}
%
%  FIGURE 20
%
% 
\begin{figure}
\plotfiddle{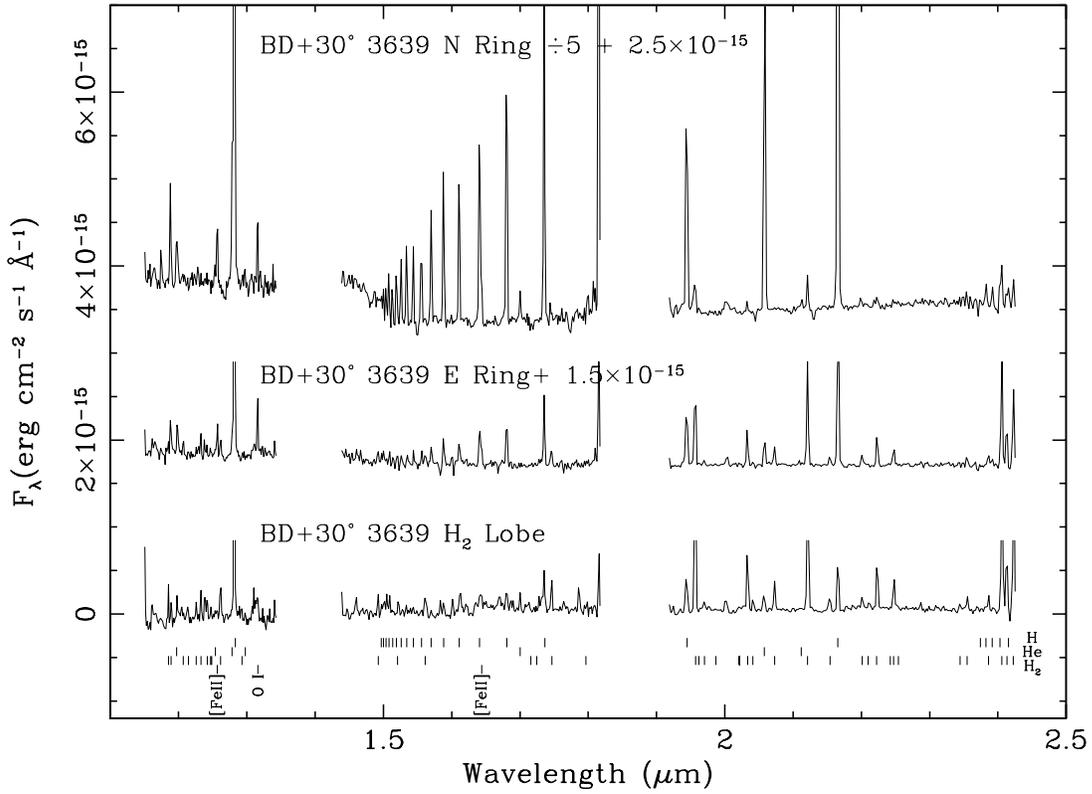}{4in}{270}{58}{58}{-250}{320}
\caption{BD+30$^{\circ}$3639 (see caption to Figure 15).}
\end{figure}
%
%  FIGURE 21
%
% 
\begin{figure}
\plotfiddle{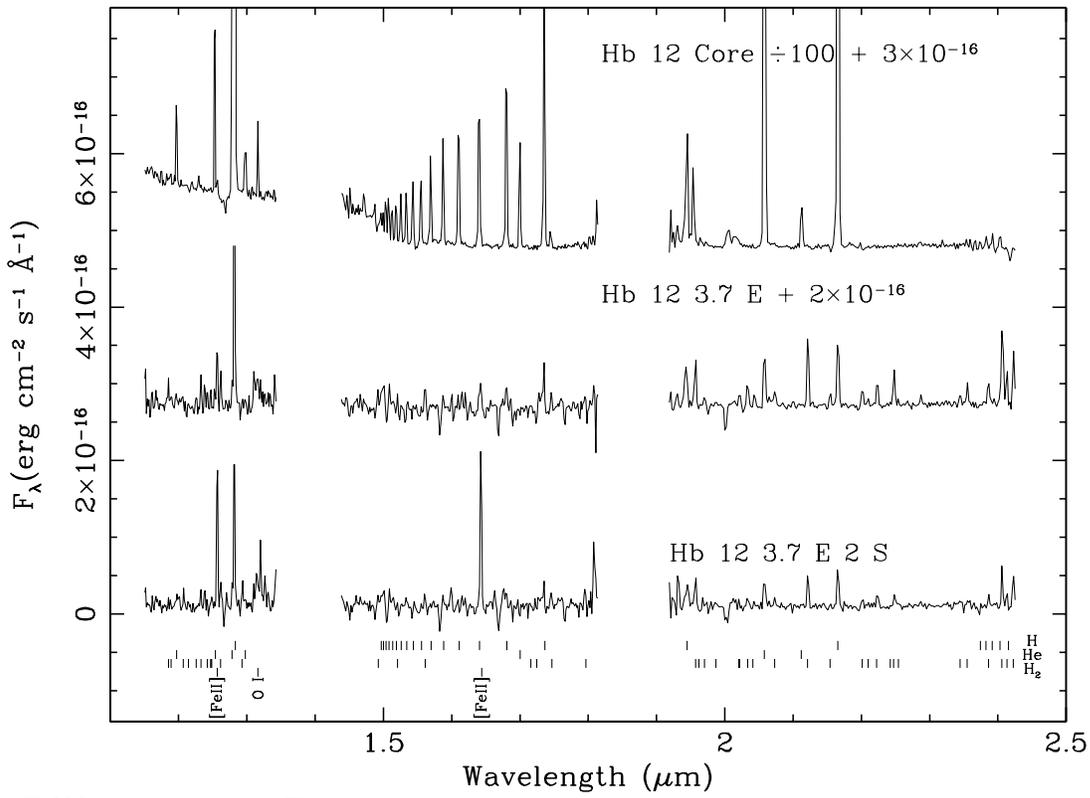}{4in}{270}{58}{58}{-250}{320}
\caption{Hubble 12 (see caption to Figure 15).}
\end{figure}
%
%  FIGURE 22
%
% 
\begin{figure}
\plotfiddle{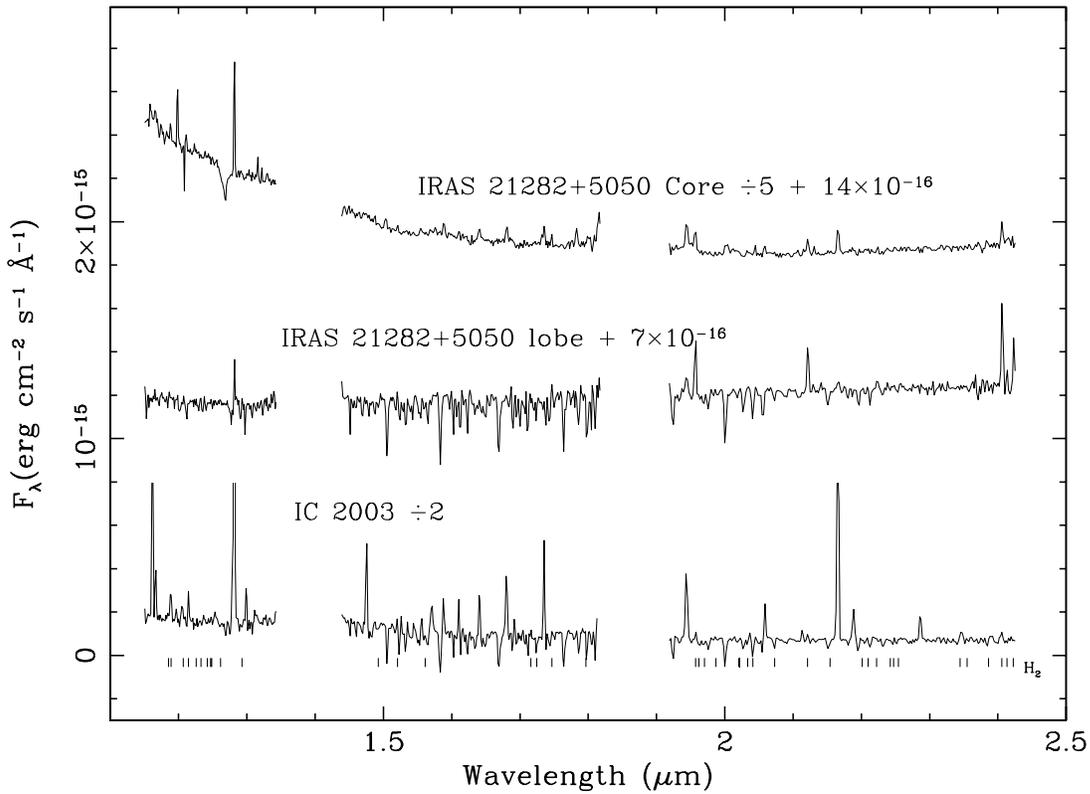}{4in}{270}{58}{58}{-250}{320}
\caption{IC 2003 and IRAS 21282+5050 (see caption to Figure 15).}
\end{figure}
%
%  FIGURE 23
%
% 
\begin{figure}
\plotfiddle{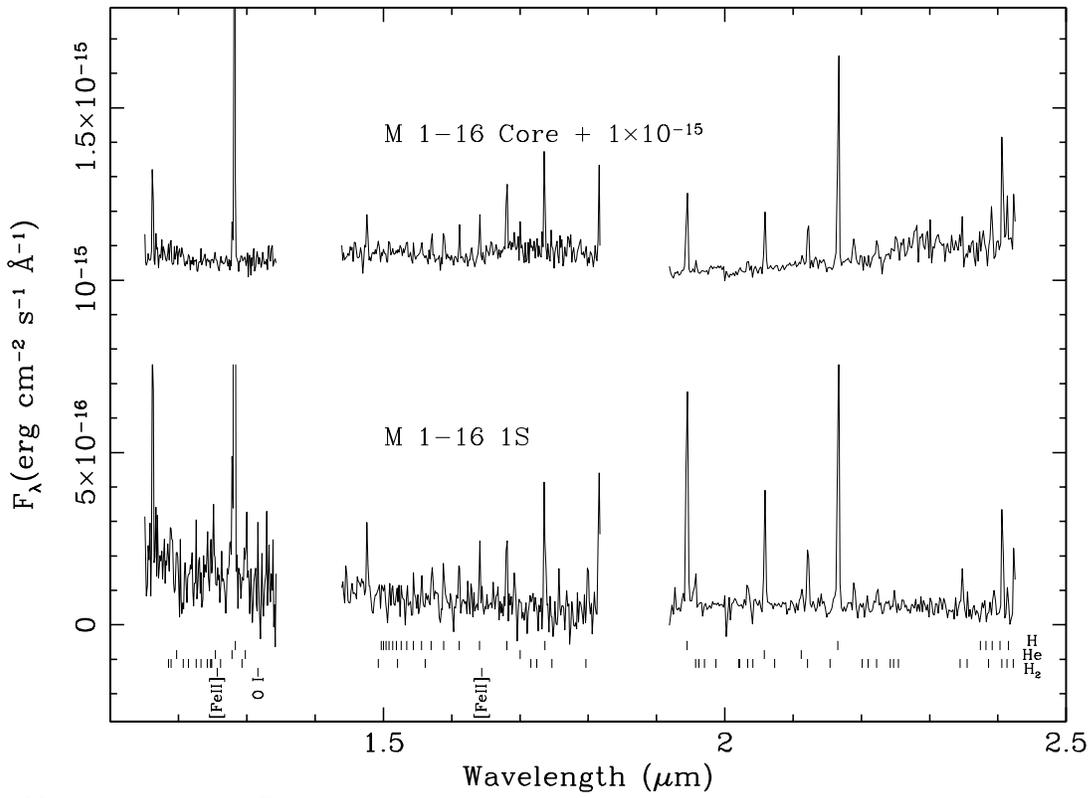}{4in}{270}{58}{58}{-250}{320}
\caption{M 1--16 (see caption to Figure 15).}
\end{figure}
%
%  FIGURE 24
%
% 
\begin{figure}
\plotfiddle{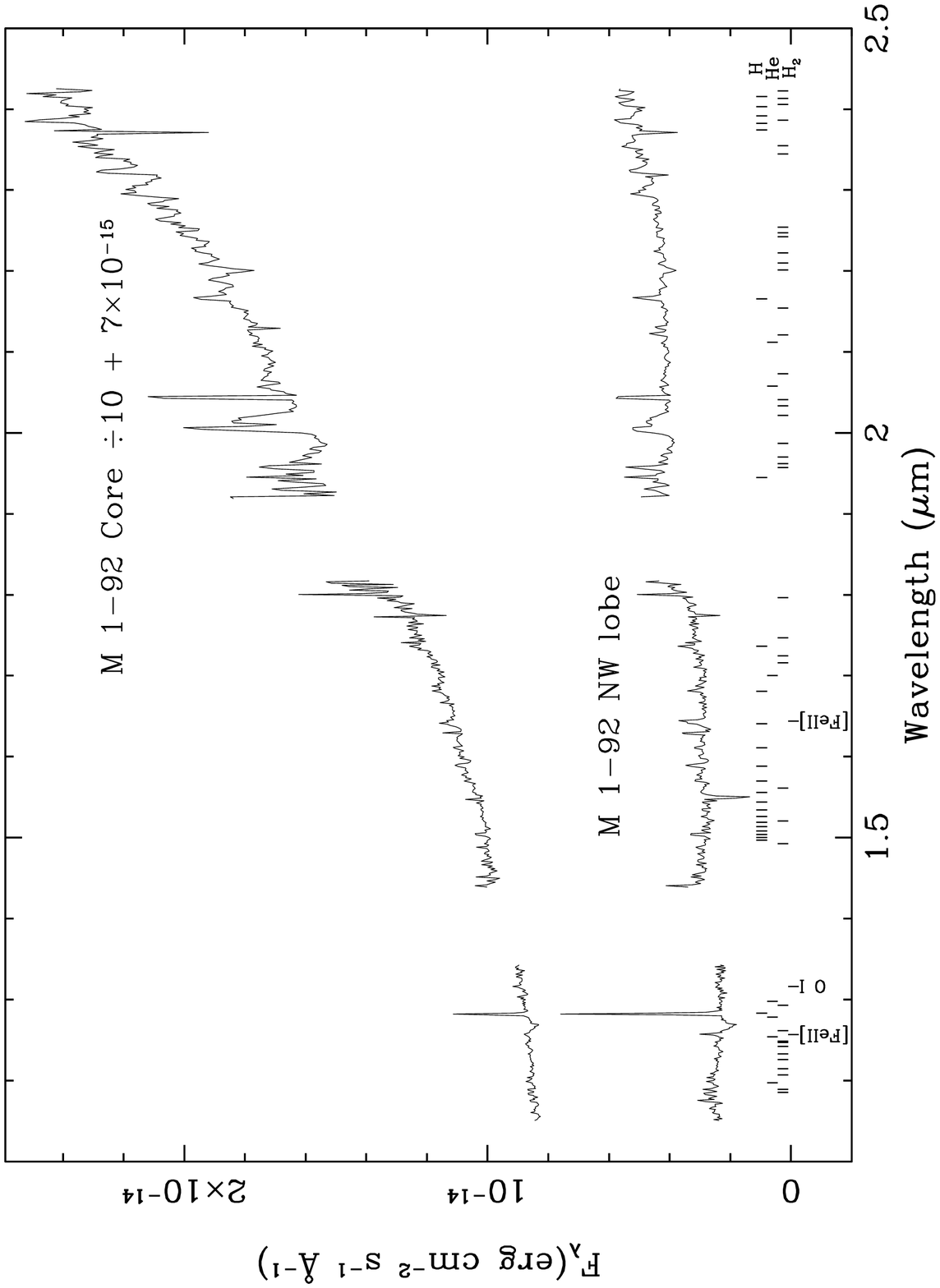}{4in}{270}{58}{58}{-250}{320}
\caption{M 1--92 (see caption to Figure 15).}
\end{figure}
%
%  FIGURE 25
%
% 
\begin{figure}
\plotfiddle{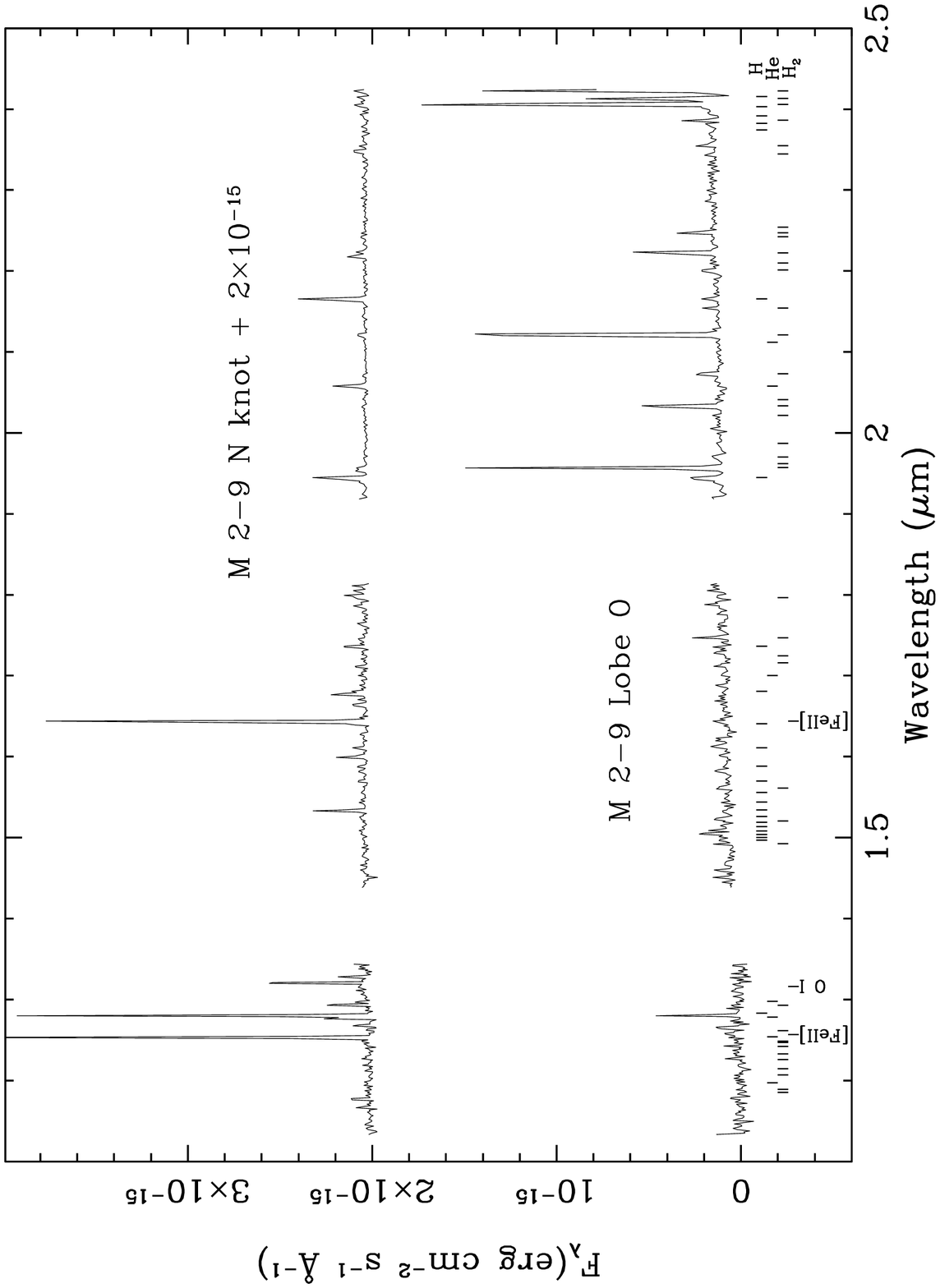}{4in}{270}{58}{58}{-250}{320}
\caption{M 2--9 (see caption to Figure 15).}
\end{figure}

%
%  FIGURE 26
%
% 
\begin{figure}
\plotfiddle{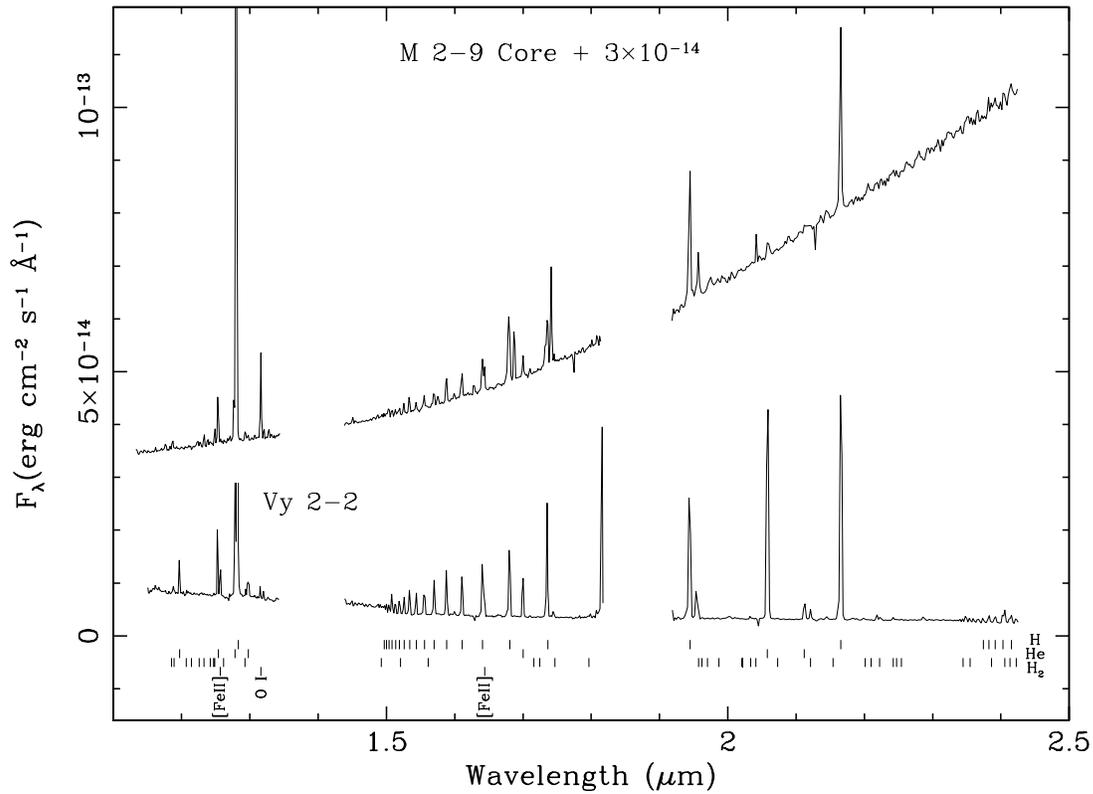}{4in}{270}{58}{58}{-250}{320}
\caption{M 2--9 and Vy 2--2 (see caption to Figure 15).}
\end{figure}
%
%  FIGURE 27
%
% 
\begin{figure}
\plotfiddle{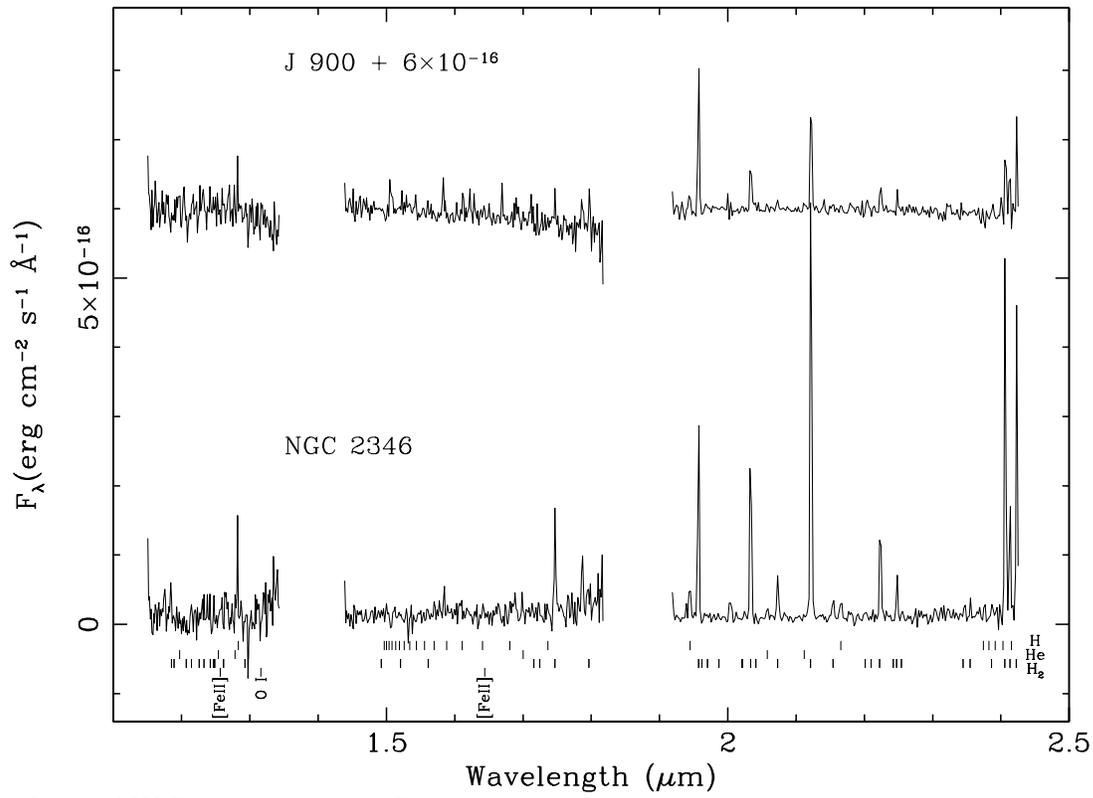}{4in}{270}{58}{58}{-250}{320}
\caption{J 900 and NGC 2346 (see caption to Figure 15).}
\end{figure}
%
%  FIGURE 28
%
% 
\begin{figure}
\plotfiddle{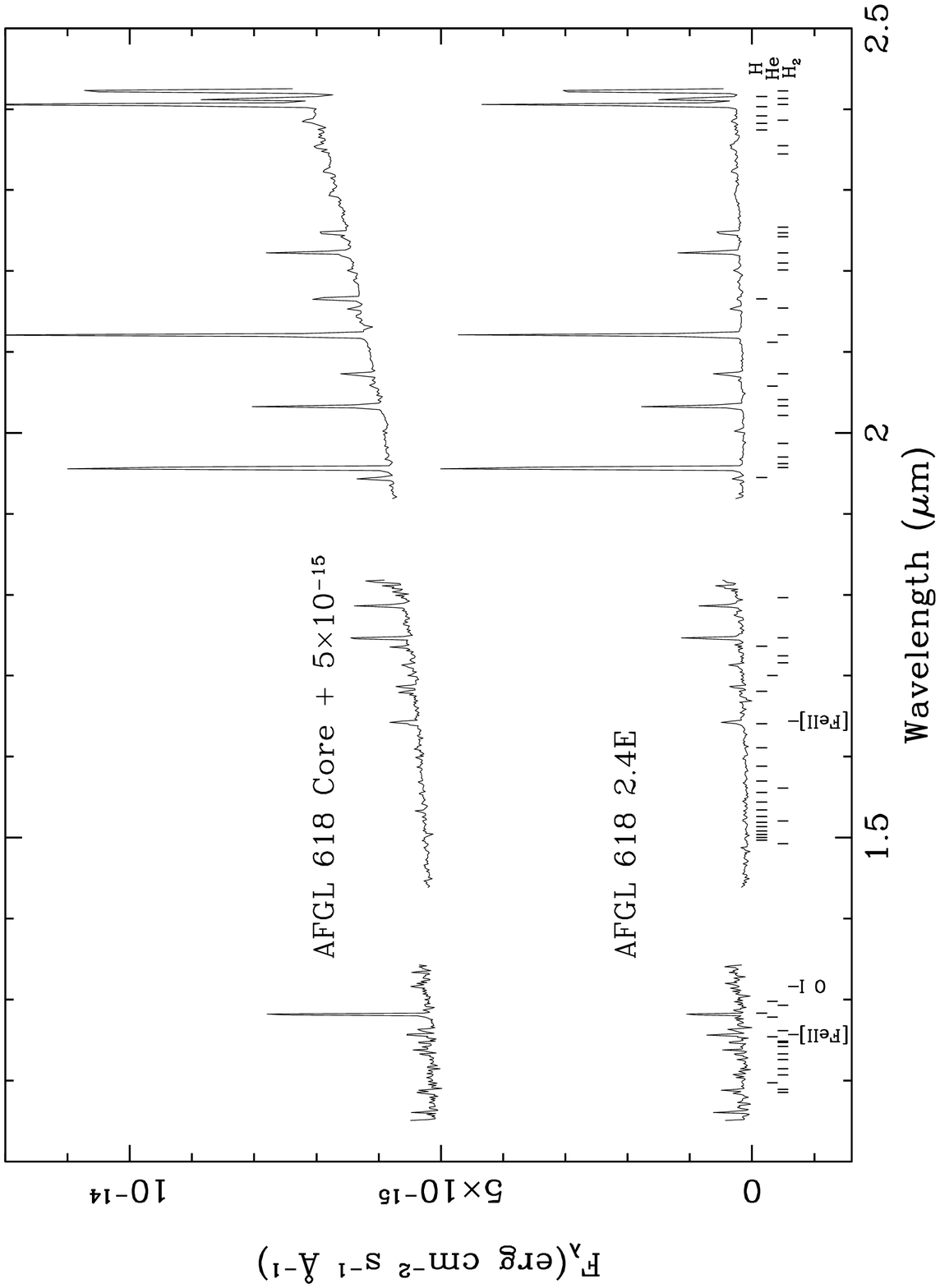}{4in}{270}{58}{58}{-250}{320}
\caption{AFGL 618 (see caption to Figure 15).}
\end{figure}
\clearpage
%
%  FIGURE 29
%
% 
\begin{figure}
\plotfiddle{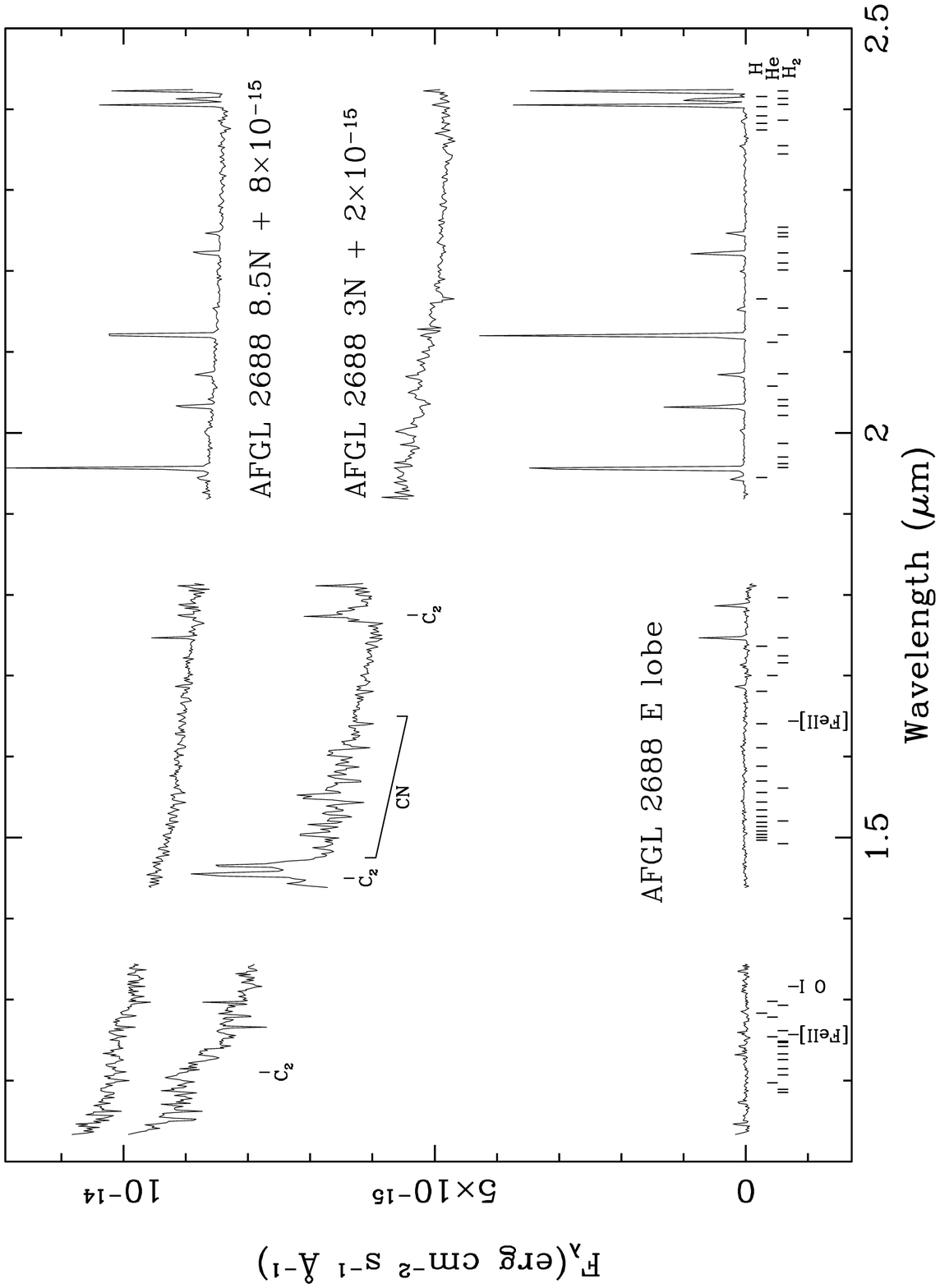}{4in}{270}{58}{58}{-250}{320}
\caption{AFGL 2688 (see caption to Figure 15).}
\end{figure}
%
%  FIGURE 30
%
% 
\begin{figure}
\plotfiddle{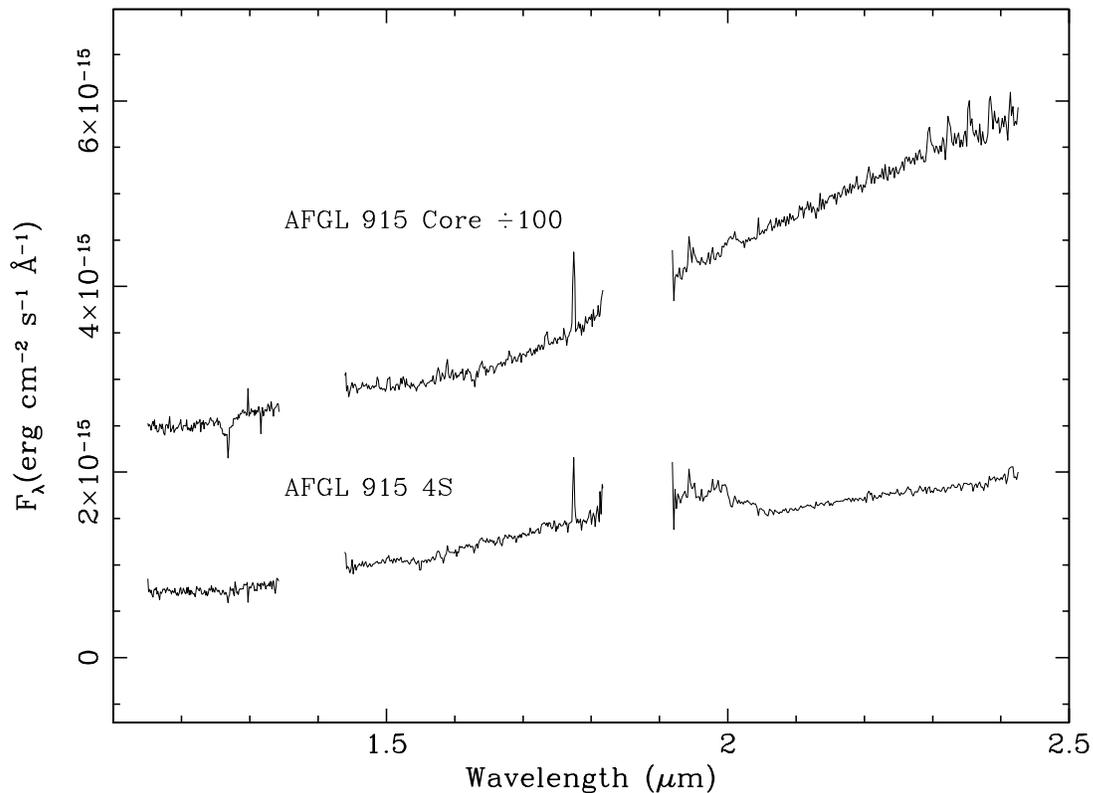}{4in}{270}{58}{58}{-250}{320}
\caption{AFGL 915.   The spectra have been offset
and scaled by the constants shown in the plot labels.  
The spectral data
points between $\lambda = 1.32 - 1.42$ \micron\ and 1.8 -- 1.9 \micron\
are in regions of poor atmospheric transmission and are not plotted.
The positions in the nebula where these spectra were taken
is given in Table 1.
}
\end{figure}
%
%  FIGURE 31
%
% 
\begin{figure}
\plotfiddle{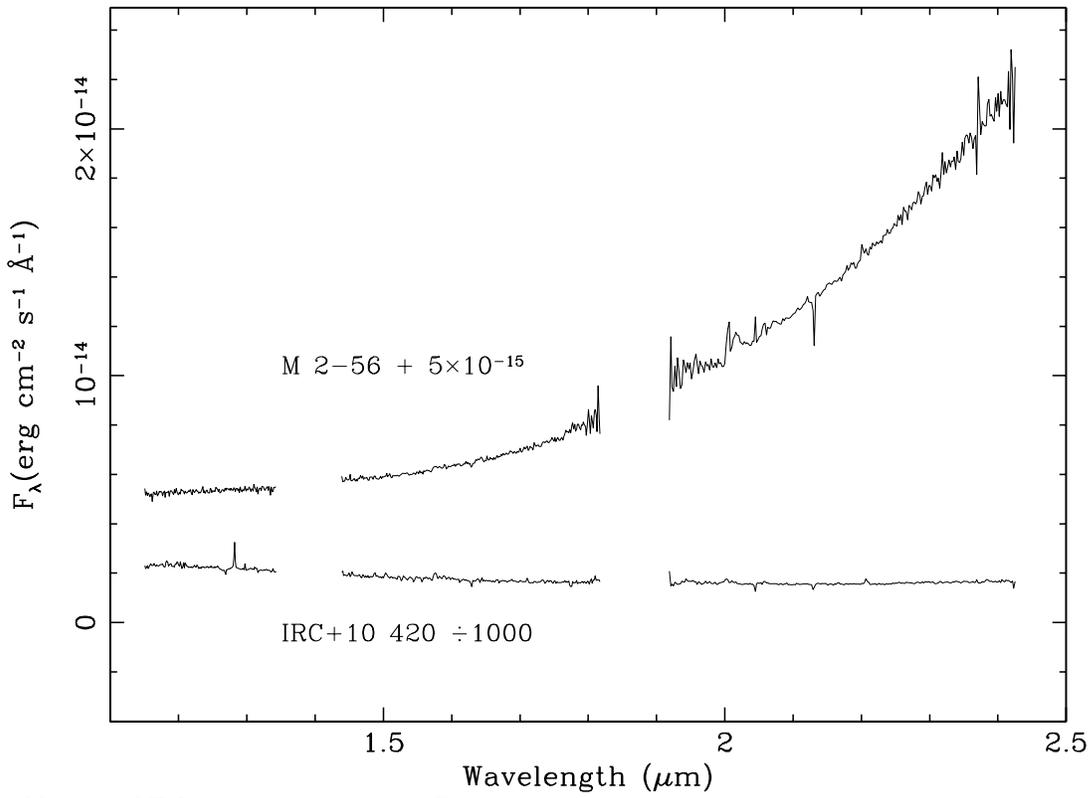}{4in}{270}{58}{58}{-250}{320}
\caption{M 2--56 and \irc10420 (see caption to Figure 30).}
\end{figure}
%
%  FIGURE 32
%
% 
\begin{figure}
\plotfiddle{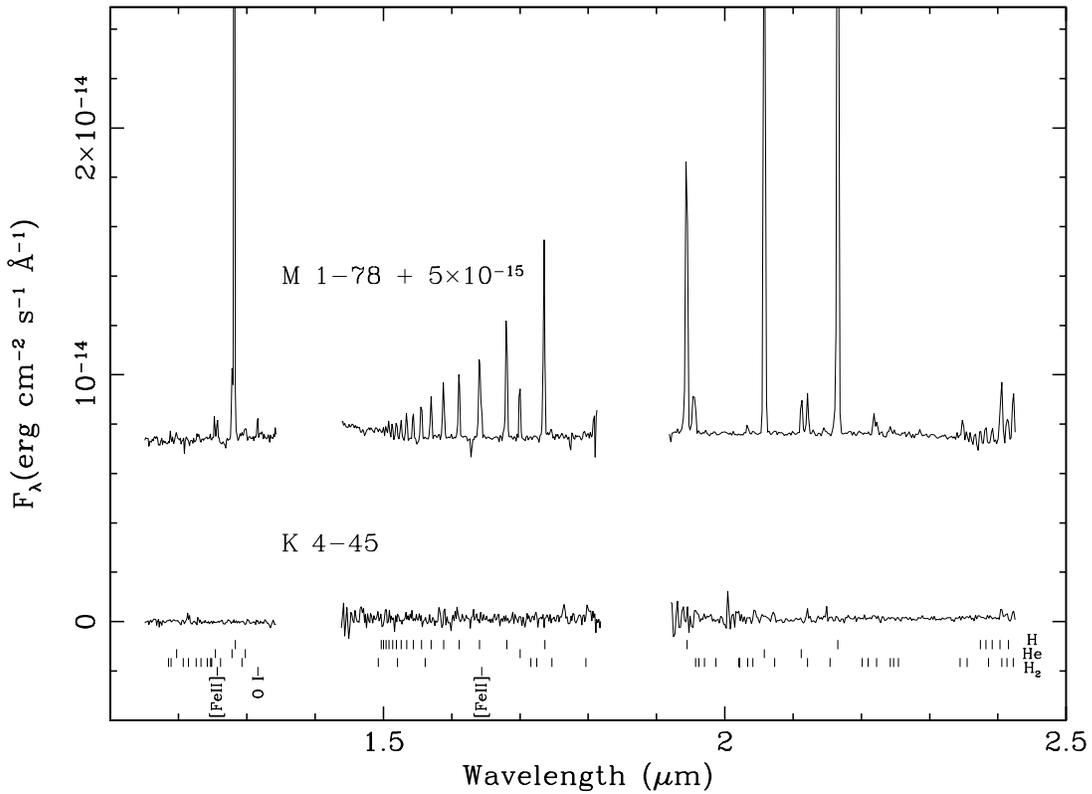}{4in}{270}{58}{58}{-250}{320}
\caption{M 1--78 and K 4--45 (see caption to Figure 15).}
\end{figure}

%
%  FIGURE 33
%
% 
\begin{figure}
\plotfiddle{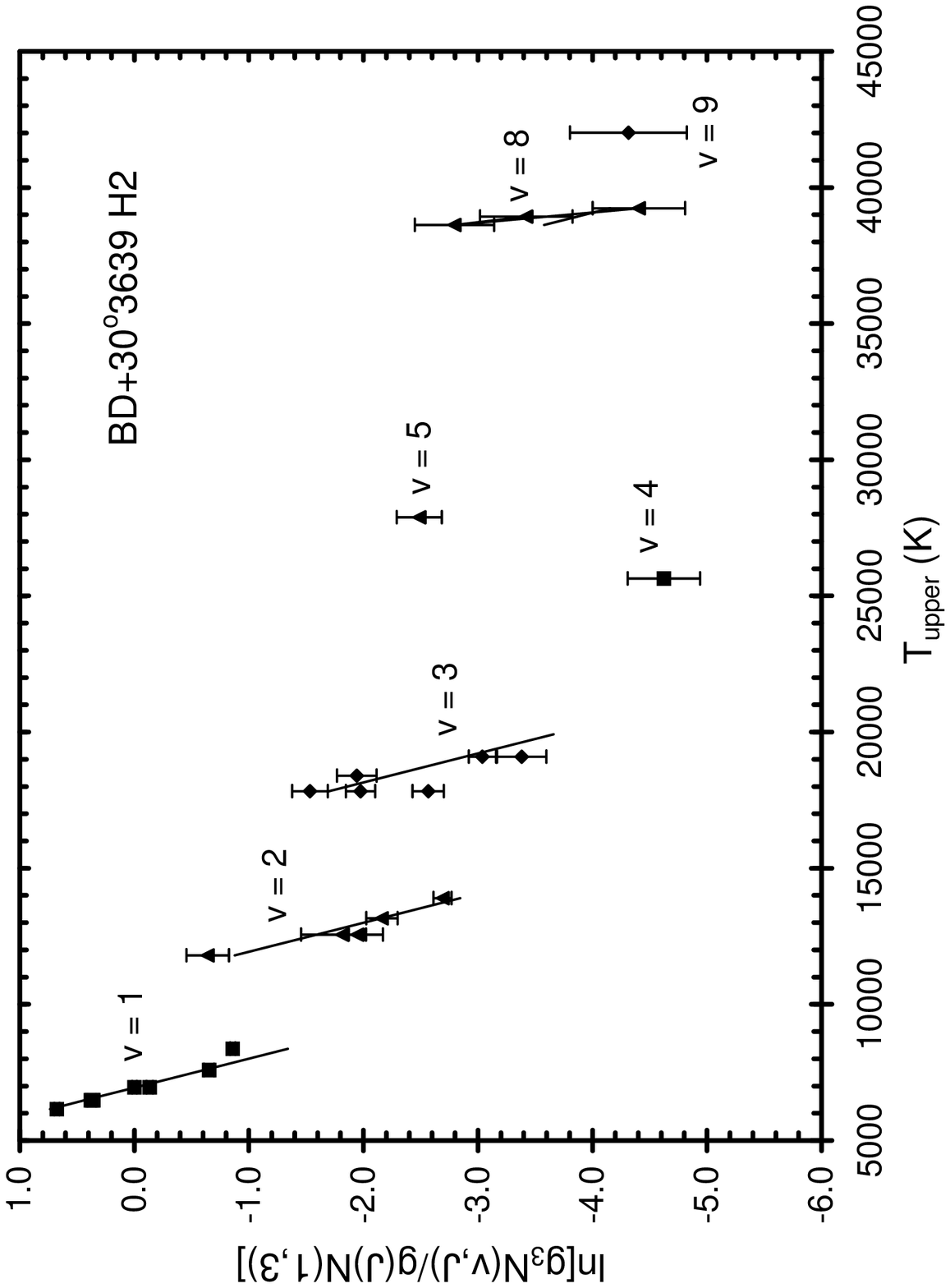}{6.5in}{270}{37}{37}{-300}{485}
\plotfiddle{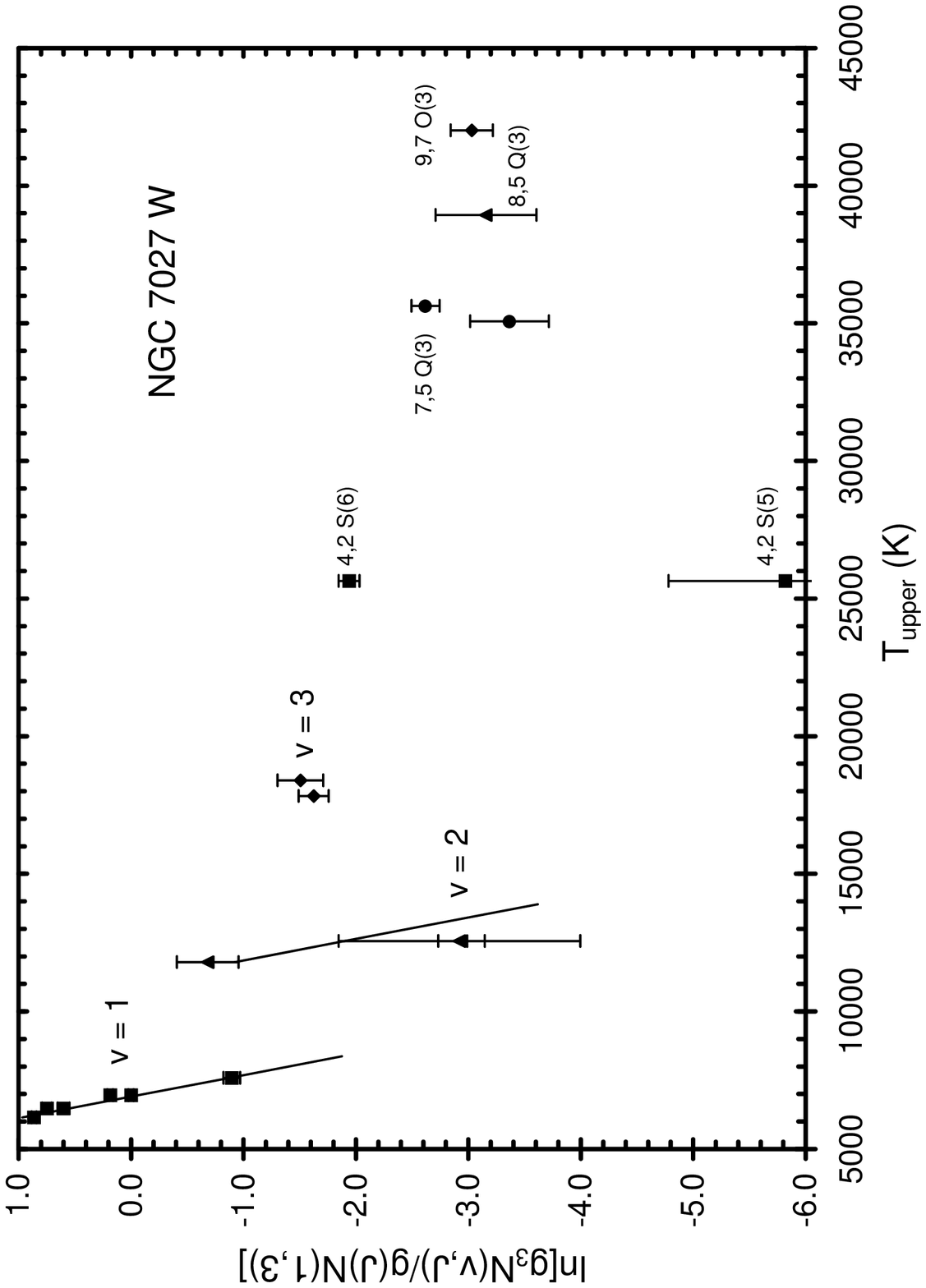}{0in}{270}{37}{37}{-20}{505}
\plotfiddle{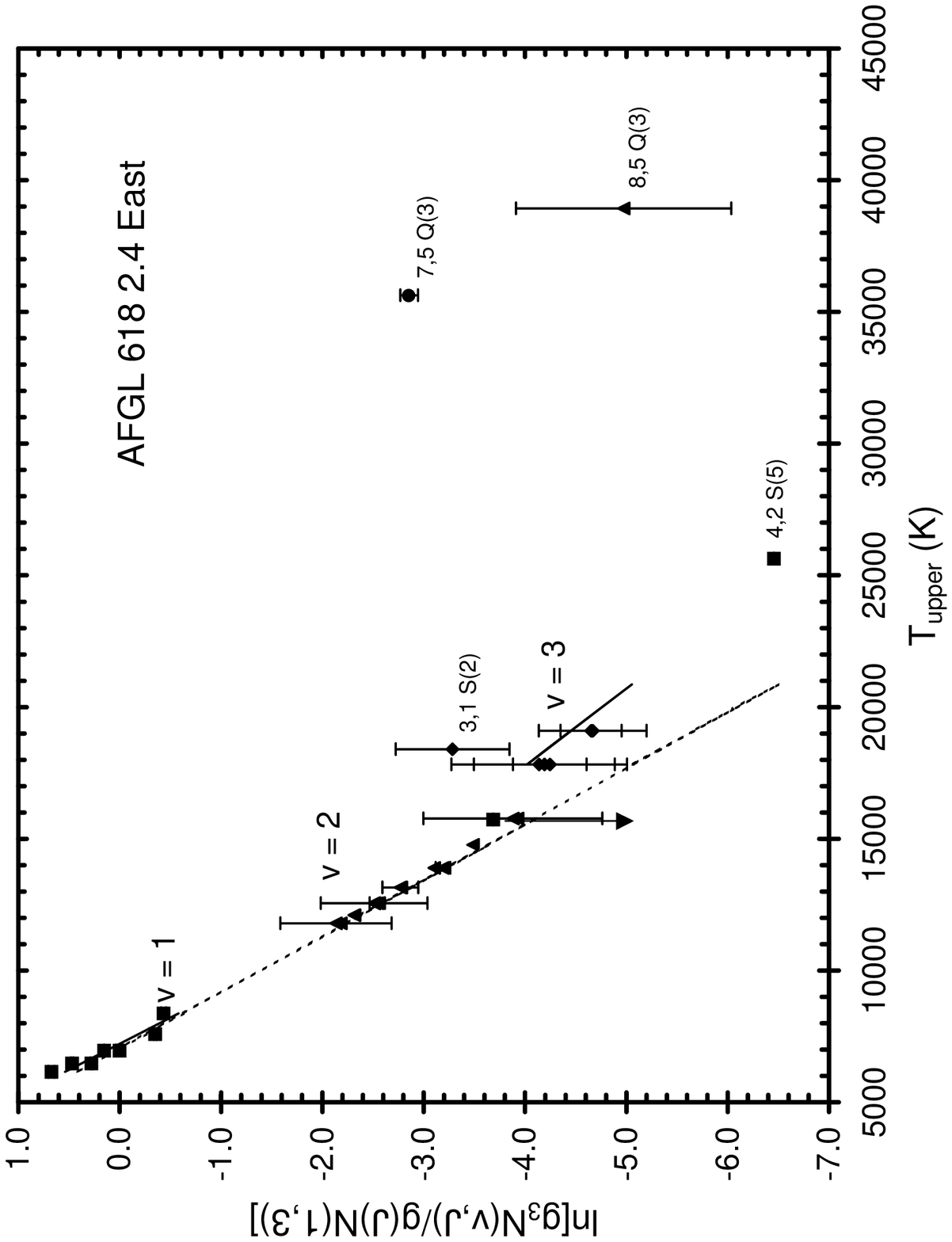}{0in}{270}{37}{37}{-300}{335}
\plotfiddle{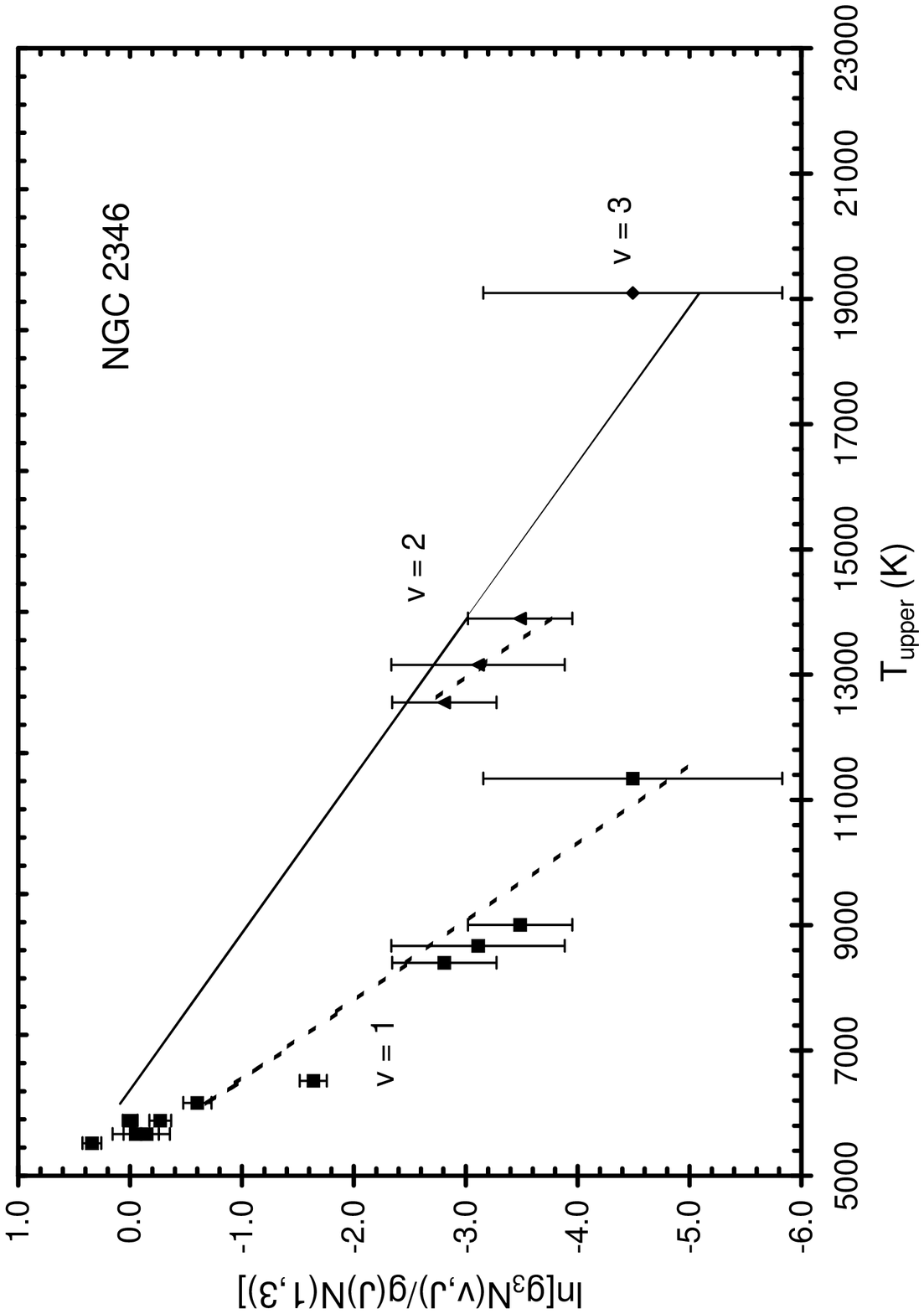}{0in}{270}{37}{37}{-20}{355}
\plotfiddle{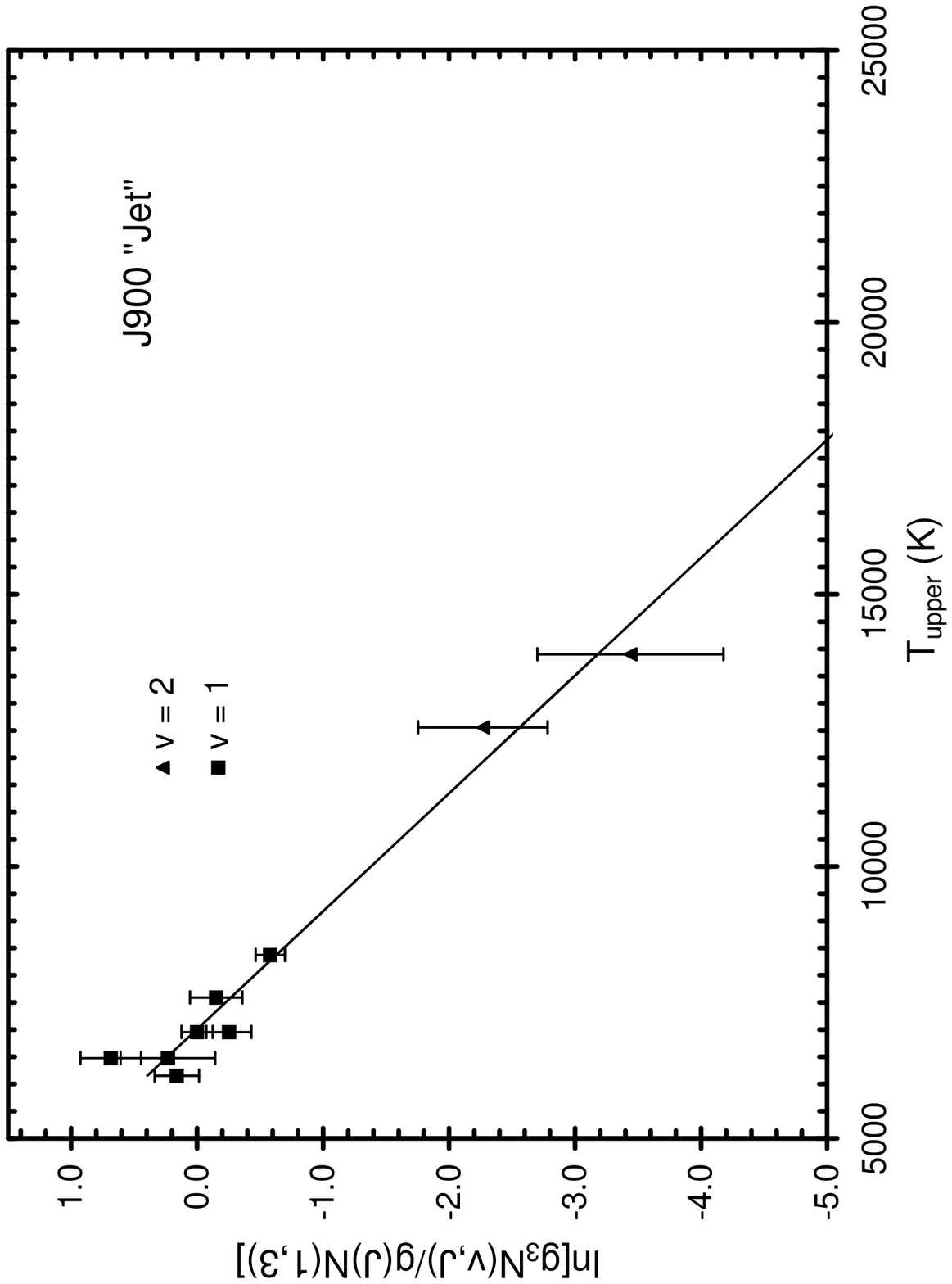}{0in}{270}{37}{37}{-300}{185}
\caption{Excitation diagram for several of the PNe detected in H$_2$.
Shown are the upper state vibration-rotation level populations
relative to that in the $v=1,\ J=3$ level plotted against the energy
of the upper state in Kelvin. The statistical weight, $g_J$, for odd
$J$ levels includes the ortho-to-para ratio as determined from the
data (Table 8), or if not determined the thermal value of 3 was
used. The data are labeled by the upper state vibration level. Linear
fits to the different vibrational levels determines the rotational
excitation temperature.  The lines shown are characteristic for this
value of $T_{ex}(J)$. Line ratios not used in the analysis because of
blending are not plotted. Dereddening of the H$_2$\ spectrum has been
done using the attenuation values shown in Table 8 (or not at all).
{\bf a)} \bd30\ at the ``H$_2$'' slit position shows strong UV excitation
(see also Shupe \etal (1998). {\bf b)} NGC 7027 at the ``W'' slit
position also shows strong UV excitation at relatively high density
(see also Graham \etal (1993a). {\bf c)} AFGL 618 at 2\farcs4 E of the
core shows a combined UV and shock-excited spectrum (see also Latter
\etal 1992). The arrow is connected to the $v = 1$ (square) point and
indicates an upper limit. The dotted line is a linear fit to the $v =
1$ and $v = 2$ data points. {\bf d)} NGC 2346 is also dominated by UV
excitation. The solid line is a linear fit ($T_{ex} = 2500$ K) to all
the data points and is offset for clarity. The dashed lines are
representative of $T_{ex}(J) = 1260$ K. {\bf e)} J 900 shows only
shock excitation of the H$_2$. For another example of a shock
(collisional) excitation diagram, see AFGL 2688 in Hora \& Latter
(1994).}
\end{figure}

\end{document}